\documentclass{article} 

\usepackage{graphicx} 
\usepackage{subfigure} 
\usepackage{times}
\usepackage{natbib}

\usepackage{algorithm}
\usepackage{algorithmic}

\usepackage{amsmath,amssymb,amsthm,bm}

\usepackage{fullpage}

\newtheorem{theorem}{Theorem}
\newtheorem{lemma}[theorem]{Lemma}
\newtheorem{proposition}[theorem]{Proposition}
\newtheorem{corollary}[theorem]{Corollary}

\newcommand\reals{\mathbb{R}}
\DeclareMathOperator*{\argmin}{argmin}
\DeclareMathOperator*{\argmax}{argmax}
\newcommand\zero{\mathbf{0}}
\DeclareMathOperator{\supp}{supp}

\newcommand\sproj[2]{H_{#1}\left( #2 \right)} 
\newcommand\compl[1]{\bar{#1}} 
\newcommand\topelem{\mathrm{top}} 

\newcommand\A{A} 
\renewcommand\b{b} 
\newcommand\I{I} 
\newcommand\curr{\I_t} 
\newcommand\currc{\compl{\I}_{t}} 

\newcommand\next{{\I_{t+1}}} 
\newcommand\nextc{\compl{\I}_{t+1}} 

\newcommand\truevec{\x^\star} 
\newcommand\true{\I^\star} 

\newcommand\lost{L_t} 
\newcommand\found{F_t} 

\newcommand\fa{FA_t} 
\newcommand\md{MD_t} 
\newcommand\co{CO_t} 

\newcommand\J{J} 
 
\newcommand\nextJ{\J_{t+1}}

\newcommand{\Acurr}{\A_{\curr}}
\newcommand{\Anext}{\A_{\next}}
\newcommand{\Atrue}{\A_{\true}}
\newcommand{\Amd}{\A_{\md}}
\newcommand{\Afound}{\A_{\found}}

\newcommand{\x}{x}
\newcommand{\y}{y}
\newcommand{\z}{z}
\newcommand{\curriter}{\x^t}
\newcommand{\nextiter}{\x^{t+1}}

\newcommand\ompr{OMPR }
\newcommand\lompr{ OMPR-Hash }
\newcommand\omprf{OMPR }
\newcommand\ihtnewton{ IHT-Newton }

\newcommand{\zmd}{\z_{\md}^{t+1}}
\newcommand{\xfa}{\x_{\fa}^t}
\newcommand{\xcurr}{\x_{\curr}^t}
\newcommand{\xtrue}{\x_{\true}^\star}
\newcommand{\xmd}{\x_{\md}^\star}
\newcommand{\xcurrco}{\x_{\co}^t}
\newcommand{\xtrueco}{\x_{\co}^\star}
\newcommand{\yfound}{\y_{\found}^{t+1}}
\newcommand{\xlost}{\x_{\lost}^t}

\renewcommand{\r}{\bm{r}}

\title{Orthogonal Matching Pursuit with Replacement}
\author{Prateek Jain\\{\tt prajain@microsoft.com} \and Ambuj Tewari\\{\tt ambuj@cs.utexas.edu} \and Inderjit S. Dhillon\\{\tt inderjit@cs.utexas.edu}}

\begin{document}

\maketitle

\begin{abstract} 
In this paper, we consider the problem of compressed sensing where the goal is to recover almost {\em all} the sparse vectors using a small number of {\em fixed} linear measurements. For this problem, we propose a novel partial hard-thresholding operator that leads to a general family of iterative algorithms. While one extreme of the family yields well known hard thresholding algorithms like ITI and HTP\cite{MalekiD10, Foucart10}, the other end of the spectrum leads to a novel algorithm that we call 
 Orthogonal Matching Pursuit with Replacement (OMPR). OMPR, like the classic greedy
algorithm OMP, adds exactly one coordinate to the support at each iteration,
based on the correlation with the current residual. However, unlike
OMP, OMPR also removes one coordinate from the support. This simple change
allows us to prove that OMPR has the best known guarantees for sparse recovery in terms of the
Restricted Isometry Property (a condition on the measurement matrix). In
contrast, OMP is known to have very weak performance guarantees under RIP.  Given its simple structure,
we are able to extend OMPR using locality sensitive hashing to get OMPR-Hash, the first
provably sub-linear (in dimensionality) algorithm for sparse recovery. Our
proof techniques are novel and flexible enough to also permit the tightest known analysis 
of popular iterative algorithms such as CoSaMP and
Subspace Pursuit.  We provide experimental results on large problems providing
recovery for vectors of size up to million dimensions.  We demonstrate that for large-scale problems our proposed methods are more robust and faster than existing
methods.

\end{abstract} 

\section{Introduction}
\label{sec:intro}
We nowadays routinely face high-dimensional datasets in diverse
application areas such as biology, astronomy, finance and the web. The
associated curse of dimensionality is often alleviated by prior knowledge that
the object being estimated has some structure. One of the most natural and
well-studied structural assumption for vectors is sparsity. Accordingly, a huge
amount of recent work in machine learning, statistics and signal processing has
been devoted to finding better ways to leverage sparse structures.
Compressed sensing, a new and active branch of modern signal processing, deals with the problem of designing
measurement matrices and recovery algorithms, such that almost {\em all} sparse signals can be recovered from a small number of measurements.
It has important applications in imaging, computer vision and machine learning (see, for example, \cite{DuarteDTLSKB08,WrightMMSHY10,HsuKLZ10}).

In this paper, we focus on the compressed sensing setting \citep{CandesT05,Donoho06} where we want to design a measurement matrix
$\A \in \reals^{m \times n}$ such that a sparse vector $\truevec \in \reals^n$ with $\|\truevec\|_0:=|\supp(\truevec)| \leq k < n$
can be efficiently recovered from the measurements $\b = \A\truevec \in \reals^{m}$. Initial work focused on various random ensembles
of matrices $\A$ such that, if $\A$ was chosen randomly from that ensemble, one would be able to recover all or almost
all sparse vectors $\truevec$ from $\A\truevec$. \cite{CandesT05} isolated a key property called the restricted 
Isometry property (RIP) and proved that, as long as the measurement matrix $\A$ satisfies RIP, the true sparse vector can be obtained
by solving an $\ell_1$-optimization problem,
\[
	\min\ \|\x\|_1 \text{ s.t. } \A\x = \b\ .
\]
The above problem can be easily formulated as a linear program and is hence efficiently solvable. We recall for the reader that
a matrix $\A$ is said to satisfy RIP of order $k$ if there is some $\delta_k \in [0,1)$ such that, for all $\x$ with $\|\x\|_0 \le k$, we have
\[
	(1-\delta_k) \| \x \|^2 \le \| \A\x \|^2 \le (1+\delta_k) \| \x \|^2 \ .
\]
Several random matrix ensembles are known to satisfy $\delta_{ck} < \theta$ with high probability provided one chooses $m = O\left( \frac{c k}{\theta^2} \log\frac{n}{k} \right)$ measurements. 
\cite{Candes08} showed that $\ell_1$-minimization recovers all $k$-sparse vectors provided $\A$ satisfies $\delta_{2k} < 0.414$
although the condition has been recently improved to $\delta_{2k} < 0.473$ \citep{Foucart10b}.
Note that, in compressed sensing, the goal is to recover all, or most, $k$-sparse signals using the {\em same} measurement matrix
$\A$. Hence, weaker conditions such as restricted convexity \cite{NRWY09} studied in the statistical literature (where the aim is to recover
a {\em single} sparse vector from noisy linear measurements) typically do not suffice. In fact, if RIP is not satisfied then multiple sparse vectors $\x$ can lead to the same observation $b$, hence making recovery of the true sparse vector impossible. 

Based on its RIP guarantees, $\ell_1$-minimization can guarantee recovery using just $O(k \log(n/k))$ measurements, but it has been observed
in practice that $\ell_1$-minimization is too expensive in large scale applications~\citep{DonohoMaMo09}, for example, when the dimensionality is in the millions.
This has sparked a huge interest in iterative methods for sparse recovery. An early classic iterative method is Orthogonal Matching Pursuit (OMP)
\citep{PatiRK93, DavidMA97} that greedily chooses elements to add to the support. It is a natural, easy-to-implement and fast method but unfortunately lacks
strong theoretical guarantees. Indeed, it is known that, if run for $k$ iterations, OMP cannot uniformly recover all $k$-sparse vectors
under an RIP condition of the form $\delta_{2k} \le \theta$ \citep{Rauhut08,MoS11}. However, \cite{Zhang10} showed that OMP, if run for $30k$ iterations, recovers the optimal solution for $\delta_{31k}\leq 1/3$; a significantly more restrictive condition than the ones required by other methods like $\ell_1$-minimization. 
 
Several other iterative approaches have been proposed that include
Iterative Soft Thresholding (IST) \citep{MalekiD10}, Iterative Hard Thresholding (IHT) \citep{BlumensathD09},
Compressive Sampling Matching Pursuit (CoSaMP) \citep{NeedellT09}, Subspace Pursuit (SP) \citep{DaiM09}, Iterative Thresholding with Inversion
(ITI) \citep{Maleki09}, Hard Thresholding Pursuit (HTP) \citep{Foucart10} and many others.  
Among the family of iterative hard thresholding algorithms, following \cite{MalekiD10}, we can identify two major subfamilies: one- and two-stage
algorithms. As their names suggest, the distinction is based on the number of stages in each iteration of the algorithm. One-stage algorithms such as IHT, ITI and HTP, decide on the
choice of the next support set and then usually solve a least squares problem on the updated support.
The one-stage methods always set the support set to have size $k$, where $k$ is the target sparsity level.
On the other hand, two-stage algorithms, notable
examples being CoSaMP and SP, first {\em enlarge} the support set, solve a least squares on it, and then {\em reduce} the support set back again
to the desired size. A second least squares problem is then solved on the reduced support. These algorithms typically enlarge and reduce the support set
by $k$ or $2k$ elements. An exception is the two-stage algorithm FoBa \cite{Zhang08} that adds and removes single elements
from the support. However, it differs from our proposed methods as its analysis requires very restrictive RIP conditions ($\delta_{8k} < 0.1$ as quoted in \cite{HsuKLZ10}) and the connection to
locality sensitive hashing (see below) is not made. Another algorithm with replacement steps appears in \cite{ShwartzSZ10}. However, the algorithm
and the setting under which it is analyzed are different from ours.

In this paper, we present, and provide a unified analysis for a family of one-stage iterative hard thresholding algorithms. The family is parameterized
by a positive integer $l \le k$. At the extreme value $l = k$, we recover the algorithm ITI/HTP. At the other extreme $k=1$, we get a novel algorithm
that we call Orthogonal Matching Pursuit with Replacement (OMPR). OMPR can be thought of as a simple modification of the classic greedy algorithm OMP:
instead of simply {\em adding} an element to the existing support, it {\em replaces} an existing support element with a new one.
Surprisingly, this change allows us to prove sparse recovery under the condition $\delta_{2k} < 0.499$. This is the best $\delta_{2k}$ based
RIP condition under which {\em any} method, including $\ell_1$-minimization, can be shown to provably perform sparse recovery.

OMPR also lends itself to a faster implementation using locality sensitive
hashing (LSH). This allows us to provide recovery guarantees using an algorithm
whose run-time is provably sub-linear in $n$, the number of dimensions.  An
added advantage of OMPR, unlike many iterative methods, is that no careful
tuning of the step-size parameter is required even under noisy settings or even
when RIP does not hold. The default step-size of $1$ is always guaranteed to
converge to at least a local optima. 

Finally, we show that our proof techniques used in the analysis of the OMPR
family are useful in tightening the analysis of two-stage algorithms, such as
CoSaMP and SP, as well. As a result, we are able to prove better recovery guarantees for these algorithms --- $\delta_{4k} < 0.35$ for CoSaMP
and $\delta_{3k} < 0.35$ for SP. We hope that this unified analysis sheds more light on
the interrelationships between the various kinds of iterative hard thresholding
algorithms.

In summary, the contributions of this paper are as follows.
\addtolength{\topsep}{-5pt}
\addtolength{\itemsep}{-5pt}
\begin{itemize}
\item
We present a family of iterative hard thresholding algorithms that on 
one end of the spectrum includes existing algorithms such as ITI/HTP while
on the other end gives OMPR. OMPR is an improvement over the classical OMP method as it enjoys better theoretical
guarantees and is also better practically as shown in our experiments.
\item
Unlike other improvements over OMP, such as CoSaMP or SP, OMPR changes only one element of the support at a
time. This allows us to use Locality Sensitive Hashing (LSH) to speed it up
resulting in the first provably sub-linear (in the ambient dimensionality $n$)
time sparse recovery algorithm.  
\item
We provide a general proof for all the algorithms in our partial hard thresholding based family. In particular, we can guarantee recovery using OMPR, under both noiseless and noisy settings,
provided $\delta_{2k} < 0.499$. This is the least restrictive $\delta_{2k}$ condition under which {\em any} efficient sparse recovery method is known to work. Furthermore, our proof technique can be used to provide a general theorem that provides the least restrictive known guarantees for all the two-stage algorithms such as CoSamp and SP (see Appendix~\ref{sec:2stage}). 
 \end{itemize}

All proofs omitted from the main body of the paper can be found in the appendix.


\section{Orthogonal Matching Pursuit with Replacement}
\label{sec:ompr} 
Orthogonal matching pursuit (OMP), is a classic iterative algorithm for sparse recovery.
At every stage, it selects a coordinate to include in the current support set by maximizing the inner product between
columns of the measurement matrix $\A$ and the current residual $\b - \A x^t$. Once the new coordinate has been added, it
solves a least squares problem to fully minimize the error on the current support set. As a result, the residual becomes orthogonal
to the columns of $\A$ that correspond to the current support set. Thus, the least squares step is also referred to as {\em orthogonalization}
by some authors \citep{DavenportW10}.

Let us briefly explain some of our notation. We use the MATLAB notation:
\[
	\A \backslash \b := \argmin_{\x} \| \A \x - \b \|_2 \ .
\]
The hard thresholding operator $H_k(\cdot)$ sorts its argument vector in decreasing order (in absolute value) and retains only the
top $k$ entries. It is defined formally in the next section. Also, we use subscripts to denote sub-vectors and submatrices, e.g. if $I\subseteq[n]$
is a set of cardinality $k$ and $\x \in \reals^n$, $\x_I \in \reals^k$ denotes the sub-vector of $\x$ indexed by $I$. Similarly, $\A_I$
for a matrix $\A \in \reals^{m \times n}$ denotes a sub-matrix of size $m \times k$ with columns indexed by $I$. The complement of set $I$ is denoted by $\bar{I}$ and $\x_{\bar{I}}$ denotes the subvector not indexed by $I$. The support (indices of non-zero entries) of a vector $\x$ is denoted by $\supp(\x)$.

Our new algorithm called Orthogonal Matching Pursuit with Replacement (\ompr), shown as Algorithm~\ref{alg:ompr}, differs from
OMP in two respects. First, the selection of the coordinate to include is based not just on the magnitude of entries in $\A^T(\b - \A \x^t)$
but instead on a weighted combination $\x^t + \eta \A^T(\b - \A \x^t)$ with the step-size $\eta$ controlling the relative importance of the
two addends. Second, the selected coordinate {\em replaces} one of the existing elements in the support, namely the one corresponding to the
minimum magnitude entry in the weighted combination mentioned above.

Once the support $\next$ of the next iterate has been determined, the actual iterate $\nextiter$ is obtained by solving the least squares problem:
\[
	\nextiter = \argmin_{\x\::\:\supp(\x)=\next}\ \| \A\x - \b \|_2 \ .
\]
Note that if the matrix $\A$ satisfies RIP of order $k$ or larger, the above problem will be well conditioned and can be solved quickly and reliably using
an iterative least squares solver.
\begin{figure*}[tb]
\begin{minipage}{.5\textwidth}
 \begin{algorithm}[H]
	\caption{\ompr}
	\label{alg:ompr}
	\begin{algorithmic}[1]
	\STATE {\bfseries Input:} matrix $\A$, vector $\b$, sparsity level $k$
	\STATE {\bfseries Parameter:} step size $\eta > 0$
	\STATE Initialize $\x^1$ s.t. $|\supp(\x^1)| = k$
	\FOR{$t=1$ {\bfseries to} $T$}
                \STATE $\z^{t+1}\gets \x^t + \eta \A^T(\b - \A \x^t)$
		\STATE $j_{t+1} \gets \argmax_{j \notin \curr} |\z^{t+1}_j|$
		\STATE $\nextJ \gets \curr \cup \{ j_{t+1} \}$
		\STATE $\y^{t+1} \gets \sproj{k}{\z^{t+1}_{\nextJ}}$
		\STATE $\next \gets \supp(\y^{t+1})$
		\STATE $\nextiter_\next \gets \Anext \backslash \b,\ \nextiter_{\nextc} \gets \zero$
	\ENDFOR
	\end{algorithmic}
\end{algorithm}
\end{minipage}
\begin{minipage}{.5\textwidth}
\begin{algorithm}[H]
	\caption{\omprf($l$)}
	\label{alg:omprfamily}
	\begin{algorithmic}[1]
	\STATE {\bfseries Input:} matrix $\A$, vector $\b$, sparsity level $k$
	\STATE {\bfseries Parameter:} step size $\eta > 0$
	\STATE Initialize $\x^1$ s.t. $|\supp(\x^1)| = k$
	\FOR{$t=1$ {\bfseries to} $T$}
                \STATE $\z^{t+1}\gets \x^t + \eta \A^T(\b-\A \x^t)$
		\STATE $\topelem_{t+1} \gets$ indices of top $l$ elements of $|\z^{t+1}_{\currc}|$
		\STATE $\nextJ \gets \curr \cup \topelem_{t+1}$
		\STATE $\y^{t+1} \gets \sproj{k}{\z^{t+1}_{\nextJ}}$
		\STATE $\next \gets \supp(\y^{t+1})$
		\STATE $\nextiter_\next \gets \Anext \backslash \b,\ \nextiter_{\nextc} \gets \zero$
	\ENDFOR	
	\end{algorithmic}
\end{algorithm}
\end{minipage}
\end{figure*}
We will show that OMPR, unlike OMP,  recovers any $k$-sparse vector
under the RIP based condition $\delta_{2k} \le 0.499$. This appears to be the least restrictive recovery condition (i.e., best known condition) under which {\em any} method, be it
basis pursuit ($\ell_1$-minimization) or some iterative algorithm, is guaranteed to recover {\bf all} $k$-sparse vectors.

In the literature on sparse recovery, RIP based conditions of a different order other than $2k$ are often provided. It is seldom possible
to directly compare two conditions, say, one based on $\delta_{2k}$ and the other based on $\delta_{3k}$. Foucart \citep{Foucart10} has given
a heuristic to compare such RIP conditions based on the number of samples it takes in the Gaussian ensemble to satisfy a given RIP condition.
This heuristic says that an RIP condition of the form $\delta_{ck} < \theta$ is less restrictive if the ratio $c/\theta^2$ is smaller. For the OMPR
condition $\delta_{2k} < 0.499$, this ratio is $2/0.499^2 \approx 8$ which makes it heuristically the least restrictive RIP condition for sparse recovery.
\begin{theorem}[Noiseless Case]
\label{thm:ompr}
Suppose the vector $\truevec \in \mathbb{R}^n$ is $k$-sparse and the matrix $A$ satisfies $\delta_{2k} < 0.499$ and $\delta_2 < 0.002$. Then \ompr\ recovers
$\epsilon$ approximation to $\truevec$ from measurements $\b = \A \truevec$ in $O(k\log k/\epsilon)$ iterations.
\end{theorem}
\begin{theorem}[Noisy Case]
\label{thm:ompr_noisy}
Suppose the vector $\truevec \in \mathbb{R}^n$ is $k$-sparse and the matrix $A$ satisfies $\delta_{2k}<.499$ and $\delta_2<0.002$. Then, in O($k\log k/\epsilon$) iterations \ompr\ converges to $C+\epsilon$ approximate solution, i.e., $f(x)=1/2\|A(\x-\x^*)+e\|^2\leq \frac{C+\epsilon}{2}\|e\|^2$ from measurements $\b = \A \truevec+e$. $C>0$ is a universal constant and is dependent only on $\delta_{2k}$. 
\end{theorem}
The above theorems are actually special cases of our convergence results for a family of algorithms that contains OMPR as a special case. We now
turn our attention to this family. We note that the condition $\delta_2 < 0.002$ is very mild and will typically hold for standard random matrix ensembles
as soon as the number of rows sampled is larger than a fixed universal constant.

\section{A New Family of Iterative Algorithms}
\label{sec:family}
In this section we show that \ompr\ is one particular member of a family of
algorithms parameterized by a single integer $l \in \{ 1,\ldots,k\}$. The $l$-th member of this
family, \omprf($l$), shown in Algorithm~\ref{alg:omprfamily}, replaces at most $l$ elements of the current support with new elements. \ompr\ corresponds
to the choice $l=1$. Hence, \ompr\ and \omprf($1$) refer to the same algorithm.

Our first result in this section connects the \omprf\ family to hard thresholding. Given a set $\I$ of cardinality
$k$, define the partial hard thresholding operator
\begin{align}
\sproj{k}{\z ; I,l} &:= \argmin_{ \stackrel{\|\y\|_0\le k}{|\supp(\y)\backslash \I|\le l} }\ \|\y - \z\| \ .
\end{align}
As is clear from the definition, the operator tries to find a vector $\y$ close to a given vector $\z$ under two constraints:
(i) the vector $\y$ should have bounded support ($\|\y\|_0 \le k$), and (ii) its support should not include more than $l$ new elements
outside a given support $I$. 

The name partial hard thresholding operator is justified because of the following reasoning.
When $l=k$, the constraint $|\supp(\y)\backslash \I| \le l$ is trivially implied by $\|\y\|_0 \le k$
and hence the operator becomes independent of $I$. In fact, it becomes identical to the standard hard thresholding operator
\begin{align}
\sproj{k}{\z ; I,k} =
\sproj{k}{\z} &:= \argmin_{ \|y \|_0 \leq k}\ \|\y - \z\| \ .
\end{align}
Even though the definition of $\sproj{k}{\z}$ seems to involve searching through $\binom{n}{k}$ subsets, it can in fact be
computed efficiently by simply sorting the vector $\z$ by decreasing absolute value and retaining the top $k$ entries.

The following result shows that even the partial hard thresholding operator is easy to compute. In fact, lines 6--8 in Algorithm~\ref{alg:omprfamily}
precisely compute $\sproj{k}{\z^{t+1} ; \curr, l}$.
\begin{proposition}
Let $|\I| = k$ and $\z$ be given. Then $\y = \sproj{k}{\z;I,l}$ can be computed
using the sequence of operations
\begin{align*}
\topelem = \text{indices of top $l$ elements of } |\z_{\compl{I}}|,\quad \J = \I \cup \topelem,\quad \y = \sproj{k}{ \z_\J } \ .
\end{align*} 
\end{proposition}
The proof of this proposition is straightforward and elementary. However, using it, we can now see that the
\omprf($l$) algorithm has a simple conceptual structure. In each iteration (with current iterate
$\curriter$ having support $\curr = \supp(\curriter)$), we do the following:
\begin{enumerate}
\item
(Gradient Descent) Form $\z^{t+1} = \curriter - \eta \A^T(\A \curriter - \b)$. Note that $\A^T(\A \curriter - \b)$ is the gradient of the objective function
$\tfrac{1}{2}\|\A\x - \b\|^2$ at $\curriter$.
\item
(Partial Hard Thresholding) Form $\y^{t+1}$ by partially hard thresholding $\z^{t+1}$ using the operator $\sproj{k}{\cdot;\curr, l}$.
\item
(Least Squares) Form the next iterate $\nextiter$ by solving a least squares problem on the support $\next$ of $\y^{t+1}$.
\end{enumerate}

A nice property enjoyed by the entire \omprf\ family is guaranteed sparse recovery under RIP based conditions.
Note that the condition under which \omprf($l$) recovers sparse vectors becomes more restrictive as $l$ increases. This could be 
an artifact of our analysis, as in experiments, we do not see any degradation in recovery ability as $l$ is increased.

\begin{theorem}[Noiseless Case]
\label{thm:family}
Suppose the vector $\truevec \in \mathbb{R}^n$ is $k$-sparse. Then \omprf($l$)\ recovers
an $\epsilon$ approximation to $\truevec$ from measurements $\b = \A \truevec$ in $O(\frac{k}{l}\log(1/\epsilon))$ iterations provided we choose a step size $\eta$ that satisfies $\eta(1+\delta_{2l}) < 1$
and $\eta(1-\delta_{2k}) > 1/2$.
\end{theorem}
\begin{theorem}[Noisy Case]
\label{thm:family_noisy}
  Suppose the vector $\truevec \in \mathbb{R}^n$ is $k$-sparse. Then \omprf($l$)\ converges to a $C+\epsilon$-approximate solution, i.e., $f(x)=1/2\|\A\x-\b\|^2\leq \frac{C+\epsilon}{2}\|e\|^2$ from measurements $\b = \A \truevec+e$ in $O(\frac{k}{l}\log ((k+\|e\|^2)/\epsilon))$ iterations provided we choose a step size $\eta$ that satisfies $\eta(1+\delta_{2l})<1$ and $\delta_{2k}<1-\frac{1}{2D\eta}$, where $D=\frac{C-\sqrt{C}}{(\sqrt{C}+1)^2}$. 
\end{theorem}
\begin{proof}
Here we provide a rough sketch of the proof of Theorem~\ref{thm:family}; the complete proof is given in Appendix A. 

Our proof uses the following  crucial observation regarding
the structure of the vector $\z^{t+1} = \curriter - \eta \A^T(\A \curriter - \b) \ .$
Due to the least squares step of the previous iteration, the current residual $\A \curriter - \b$ is orthogonal to columns of $\A_{\curr}$. This means that
\begin{align}
\z^{t+1}_{\curr} = \curriter_{\curr} \ ,\quad \z^{t+1}_{\currc} = -\eta \A_{\currc}^T(\A \curriter - \b) \ .\label{eq:zt}
\end{align}
As the algorithm proceeds, elements come in and move out of the current set $\curr$. Let us give names to the
set of found and lost elements as we move from $\curr$ to $\next$:
\begin{align*}
\text{(found)}:\ \ \found = \next \backslash \curr, \qquad \text{(lost)}: \lost = \curr \backslash \next.
\end{align*}
Hence, using \eqref{eq:zt} and updates for $y_{t+1}$: $\yfound = \z^{t+1}_{\found}=-\eta \Afound^T\A(\curriter-\truevec)$, and  $\z^{t+1}_{\lost} = \xlost$.
Now let $f(\x)=1/2\|\A\x-b\|^2$, then using {\em upper} RIP and the fact that $|\supp(\y^{t+1}-\curriter)| = |\found \cup \lost| \le 2l$, we can show that (details are in the Appendix A):
\begin{align}
f(\y^{t+1})-f(\curriter)&\leq \left(\frac{1+\delta_{2l}}{2}-\frac{1}{\eta}\right)\|\yfound\|^2
+ \frac{1+\delta_{2l}}{2} \|\xlost\|^2.
\label{eq:fdiff1_1}
\end{align}
Furthermore,  since $\y^{t+1}$ is chosen based on the $k$ largest entries in $\z^{t+1}_{\J_{t+1}}$, we have: $\|\yfound\|^2 = \|\z^{t+1}_{\found}\|^2 \geq \|\z^{t+1}_{\lost}\|^2 = \|\xlost\|^2\ .$
Plugging this into \eqref{eq:fdiff1_1}, we get:
\begin{align}
f(\y^{t+1})-f(\curriter)\leq \left(1+\delta_{2l} - \frac{1}{\eta} \right)\|\yfound\|^2 \ . 
\label{eq:fdiff2_1}
\end{align}

The above expression shows that if $\eta<\frac{1}{1+\delta_{2l}}$ then our method monotonically decreases the objective function and converges to a local optimum even if RIP is not satisfied (note that upper RIP bound is independent of lower RIP bound, and  can always be satisfied by normalizing the matrix appropriately). 

However, to prove convergence to the global optimum, we need to show that at least one new element is added at each step, i.e., $|\found|\geq 1$. Furthermore, we need to show sufficient decrease, i.e, $\|\yfound\|^2\geq c\frac{l}{k}f(\x^t)$. We show both these conditions for global convergence in Lemma~\ref{claim:2_main}, whose proof is given in Appendix~\ref{app:ompr}.

Assuming Lemma~\ref{claim:2_main},  \eqref{eq:fdiff2_1} shows that at each iteration \omprf($l$) reduces the objective function value by at least a constant fraction. 
Furthermore, if $\x^0$ is chosen to have entries bounded by $1$, then $f(\x^0) \leq (1+\delta_{2k})k$.
Hence, after $O(k/l \log(k/\epsilon))$ iterations, the optimal solution $\truevec$ would be obtained within $\epsilon$ error.
\end{proof}
\begin{lemma}
\label{claim:2_main}
Let $\delta_{2k}<1-\frac{1}{2\eta}$ and $1/2<\eta<1$. 
Then assuming $f(x^t)>0$, at least one new element is found i.e. $\found \neq \emptyset$.
Furthermore, $\|\yfound\|>\frac{l}{k}cf(\x^t)$, where $c=\min(4\eta(1-\eta)^2,2(2\eta-\frac{1}{1-\delta_{2k}}))>0$ is a constant.
\end{lemma}
{\bf Special Cases}: We have already observed that the \ompr\ algorithm of the previous section is simply \omprf($1$). Also note that Theorem~\ref{thm:ompr} immediately
follows from Theorem~\ref{thm:family}. 

The algorithm at the other extreme of $l=k$ has appeared
at least three times in the recent literature: as Iterative (hard) Thresholding with Inversion (ITI) in \cite{Maleki09},
as SVP-Newton (in its matrix avatar) in \cite{JainMD10}, and as Hard Thresholding Pursuit (HTP) in \cite{Foucart10}).
Let us call it \ihtnewton as the least squares step can be viewed as a Newton step for the quadratic objective. The above general result for the \omprf\ family immediately implies that it recovers sparse vectors
as soon as the measurement matrix $\A$ satisfies $\delta_{2k} < 1/3$.
\begin{corollary}
Suppose the vector $\truevec \in \mathbb{R}^n$ is $k$-sparse and the matrix $A$ satisfies $\delta_{2k} < 1/3$. Then \ihtnewton\ recovers
$\truevec$ from measurements $\b = \A \truevec$ in $O(\log(k))$ iterations.
\end{corollary}


\section{Tighter Analysis of Two Stage Hard Thresholding Algorithms}
Recently, \cite{MalekiD10} proposed a novel family of algorithms, namely two-stage hard thresholding  algorithms. During each iteration, these algorithms add a fixed number (say $l$) of elements to the current iterate's support set. A least squares problem is solved over the larger support set and then $l$ elements with smallest magnitude are dropped to form next iterate's support set. Next iterate is then obtained by again solving the least squares over next iterate's support set. See Appendix D for a more detailed description of the algorithm. 

Using proof techniques developed for our proof of Theorem~\ref{thm:family}, we can obtain a simple proof for the entire spectrum of algorithms in the two-stage hard thresholding family. 
\begin{theorem}
Suppose the vector $\truevec \in \{-1,1\}^n$ is $k$-sparse. Then the Two-stage Hard Thresholding algorithm with replacement size $l$ recovers
$\truevec$ from measurements $\b = \A \truevec$ in $O(k)$ iterations provided: $\delta_{2k+l}\leq .35.$
\end{theorem}
Note that CoSaMP \cite{NeedellT09} and Subspace Pursuit(SP) \citep{DaiM09} are popular special cases of the two-stage family. Using our general analysis, we are able to provide significantly less restrictive RIP conditions for recovery. 
\begin{corollary}
  CoSaMP\cite{NeedellT09} recovers $k$-sparse $\truevec\in \{-1,1\}^n$ from measurements $\b = \A \truevec$ provided $\delta_{4k}\leq 0.35.$
\end{corollary}
\begin{corollary}
  Subspace Pursuit\cite{DaiM09} recovers $k$-sparse  $\truevec\in \{-1,1\}^n$ from measurements $\b = \A \truevec$ provided $\delta_{3k}\leq 0.35.$
\end{corollary}
Note that CoSaMP's analysis given by \cite{NeedellT09} requires $\delta_{4k}\leq \mathbf{0.1}$ while Subspace Pursuit's analysis given by \cite{DaiM09} requires $\delta_{3k}\leq\mathbf{0.205}$. See Appendix D in the supplementary material for proofs of the above theorem and corollaries. 

\section{Fast Implementation Using Hashing}
\label{sec:lsh}
In this section, we discuss a fast implementation of the \ompr method using locality-sensitive hashing. The main intuition behind our approach is that the \ompr method selects at most one element at each step (given by $\argmax_i |\A_i^T(\A\x^t-\b)|$); hence, selection of the top most element is equivalent to finding the column $\A_i$ that is most ``similar'' (in magnitude) to $\r_t=\A\x^t-\b$, i.e., this may be viewed as the similarity search task for queries of the form $\r_t$ and $-\r_t$. 

To this end, we use locality sensitive hashing (LSH) ~\citep{GionisIM99}, a well known data-structure for approximate nearest-neighbor retrieval. Note that while LSH is designed for nearest neighbor search (in terms of Euclidean distances) and in general might not have any guarantees for the similar neighbor search task, we are still able to apply it to our task because we can lower-bound the similarity of the most similar neighbor. 

We first briefly describe the LSH scheme that we use. LSH generates hash bits for a vector using randomized hash functions that have the property that the probability of collision between two vectors is proportional to the similarity between them. For our problem, we use the following hash function: $h_u(x)=\text{sign}(u^Tx)$, where $u\sim N(0,I)$ is a random hyper-plane generated from the standard multivariate Gaussian distribution. It can be shown that \cite{GW94}
$$Pr[h_u(\x_1)= h_u(\x_2)]=1-\frac{1}{\pi}\cos^{-1}\left(\frac{\x_1^T\x_2}{\|\x_1\| \|\x_2\|}\right).$$
Now, an $s$-bit hash key is created by randomly sampling hash functions $h_u$, i.e., $g(x)=[h_{u_1}(x), h_{u_2}(x), \dots, h_{u_s}(x)]$, where each $u_i$ is sampled randomly from the standard multivariate Gaussian distribution. Next, $q$ hash tables are constructed during the pre-processing stage using independently constructed hash key functions $g_1, g_2, \dots, g_q$. During the query stage, a query is indexed into each hash table using hash-key functions $g_1, g_2, \dots, g_q$ and then the nearest neighbors are retrieved by doing an exhaustive search over the indexed elements. 

Below we state the following theorem from \cite{GionisIM99} that guarantees sub-linear time nearest neighbor retrieval for LSH. 
\begin{theorem}
  Let $s=O(\log n)$ and  $q=O(\log 1/\delta)n^{\frac{1}{1+\epsilon}}$, then with probability $1-\delta$, LSH recovers $(1+\epsilon)$-nearest neighbors, i.e., 
$\|\x'-\r\|^2\leq (1+\epsilon)\|\x^*-\r\|^2,$ where $\x^*$ is the nearest neighbor to $\r$ and $\x'$ is a point retrieved by LSH.
\label{thm:lsh}
\end{theorem}

However, we cannot directly use the above theorem to guarantee convergence of our hashing based \ompr algorithm as our algorithm requires finding the most similar point in terms of magnitude of the inner product. Below, we provide appropriate settings of the LSH  parameters  to guarantee sub-linear time convergence of our method under a slightly weaker condition on the RIP constant. 
A detailed proof of the theorem below can be found in Appendix~\ref{app:lsh}. 

\begin{theorem}
  \label{thm:lsh_ompr}
Let $\delta_{2k}< 1/4-\gamma$ and $\eta=1-\gamma$, where $\gamma>0$ is a small constant, then with probability $1-\delta$, \ompr with hashing converges to the optimal solution in $O(kmn^{1/(1+\Omega(1/k))}\log k/\delta)$ computational steps. 
\end{theorem}

The above theorem shows that the time complexity is sub-linear in $n$. However, currently our guarantees are not particularly strong as for large $k$ the exponent of $n$ will be close to $1$. We believe that the exponent can be improved by more careful analysis and our empirical results indicate that LSH does speed up the \ompr method significantly. 

\section{Experimental Results}
In this section we present empirical results to demonstrate accurate and fast recovery by our \ompr method. In the first set of experiments, we present phase transition diagram for \ompr and compare it to the phase transition diagram of OMP and IHT-Newton with step size $1$. For the second set of experiments, we demonstrate robustness of \ompr compared to many existing methods when measurements are noisy or smaller in number than what is required for exact recovery. For the third set of experiments, we demonstrate efficiency of our LSH based implementation by comparing recovery error and time required for our method with OMP and IHT-Newton (with step-size $1$ and $1/2$). We do not present results for the $\ell_1$/basis pursuit methods, as it has already been shown in several recent papers \citep{Foucart10,MalekiD10} that the $\ell_1$ relaxation based methods are relatively inefficient for very large scale recovery problems.

In all the experiments we generate the measurement matrix by sampling each entry independently from the standard normal distribution ${\cal N}(0, 1)$ and then normalize each column to have unit norm. The underlying $k$-sparse vectors are generated by randomly selecting a support set of size $k$ and then each entry in the support set is sampled uniformly from $\{+1,-1\}$. We use our own optimized implementation of OMP and IHT-Newton. All the methods are implemented in MATLAB and our hashing routine uses mex files. 
\begin{figure*}[t]
  \centering
  \begin{tabular}{ccc}
    \includegraphics[width=.33\textwidth]{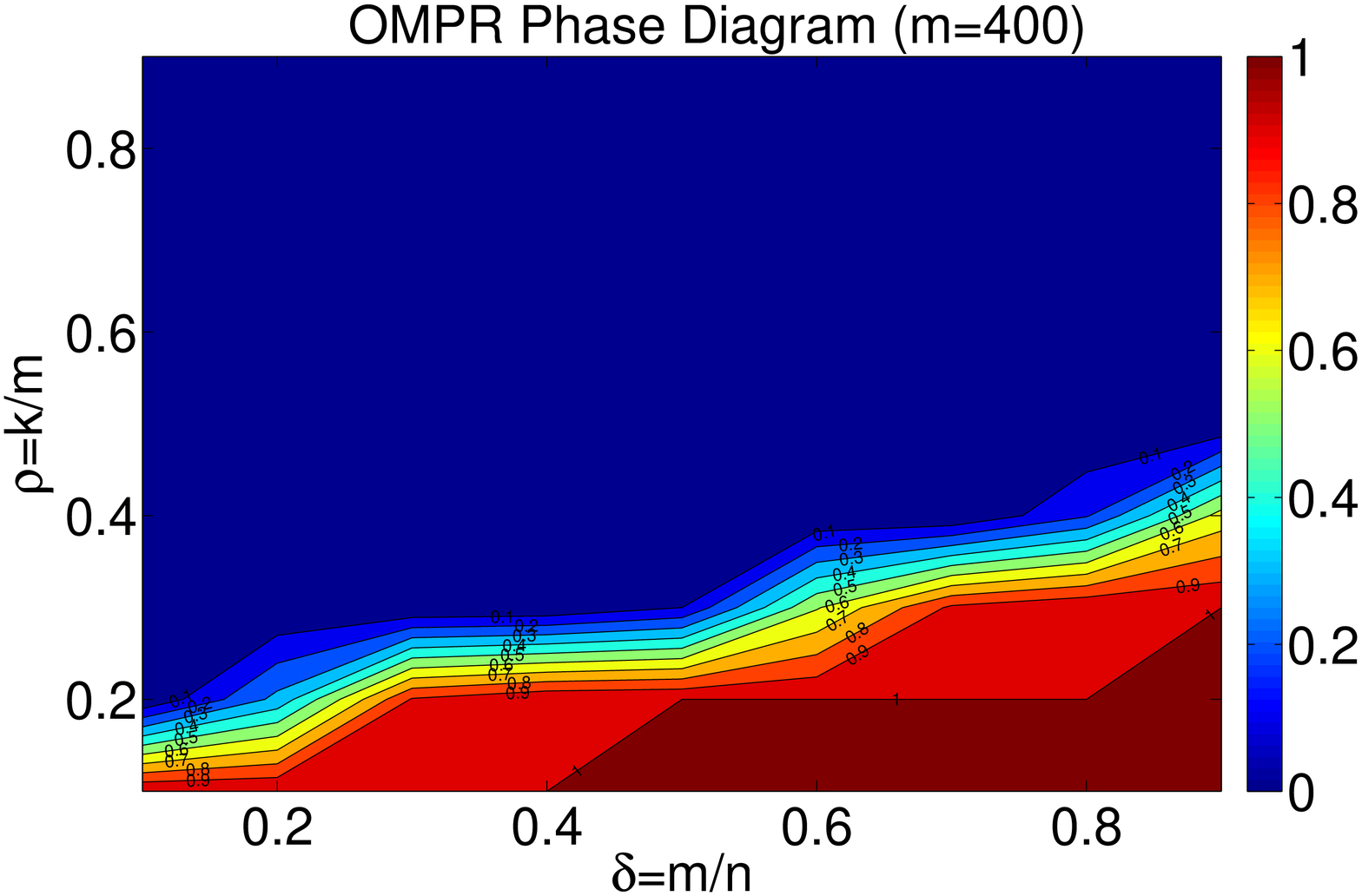}&
    \includegraphics[width=.33\textwidth]{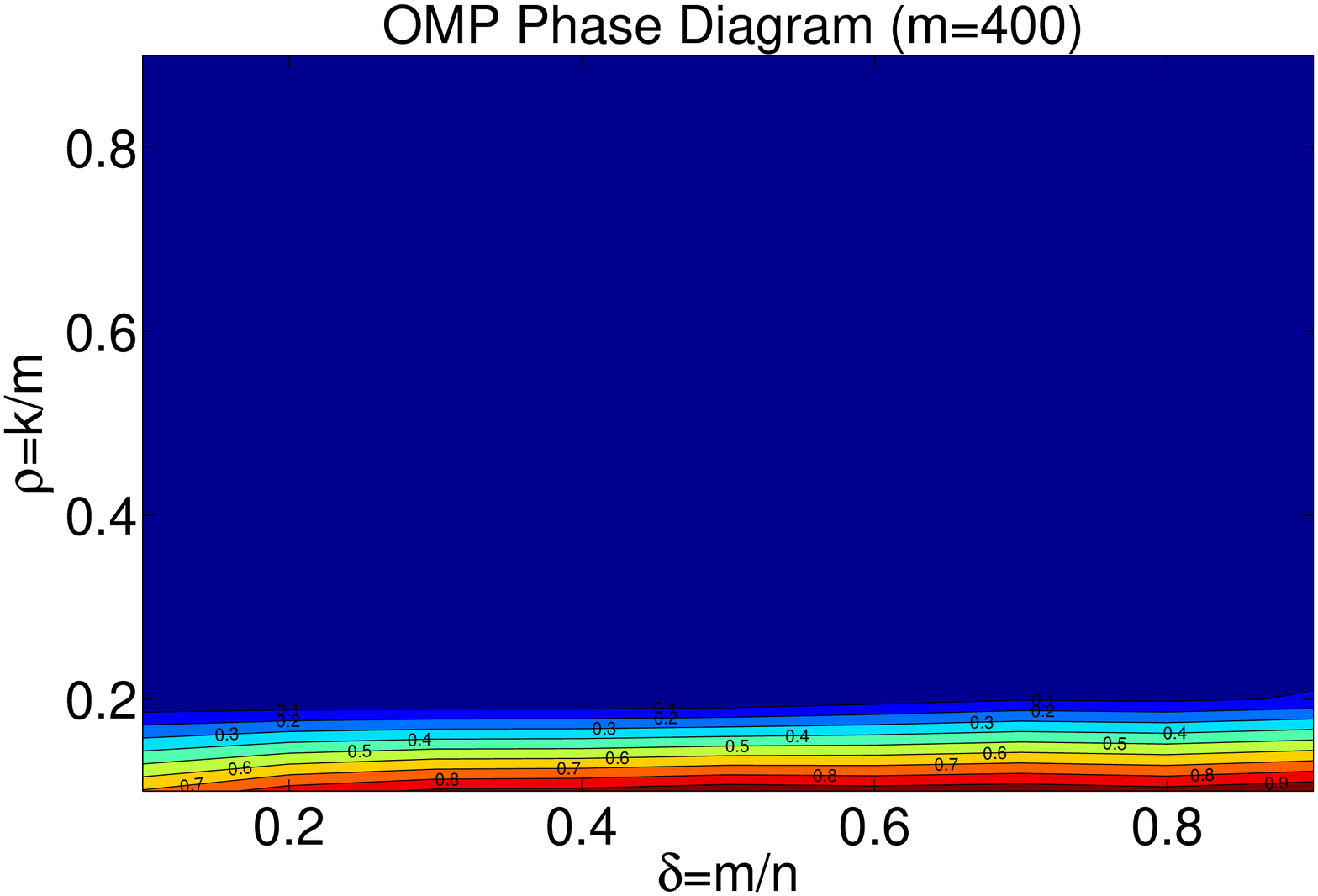}&
    \hspace*{-10pt}\includegraphics[width=.33\textwidth]{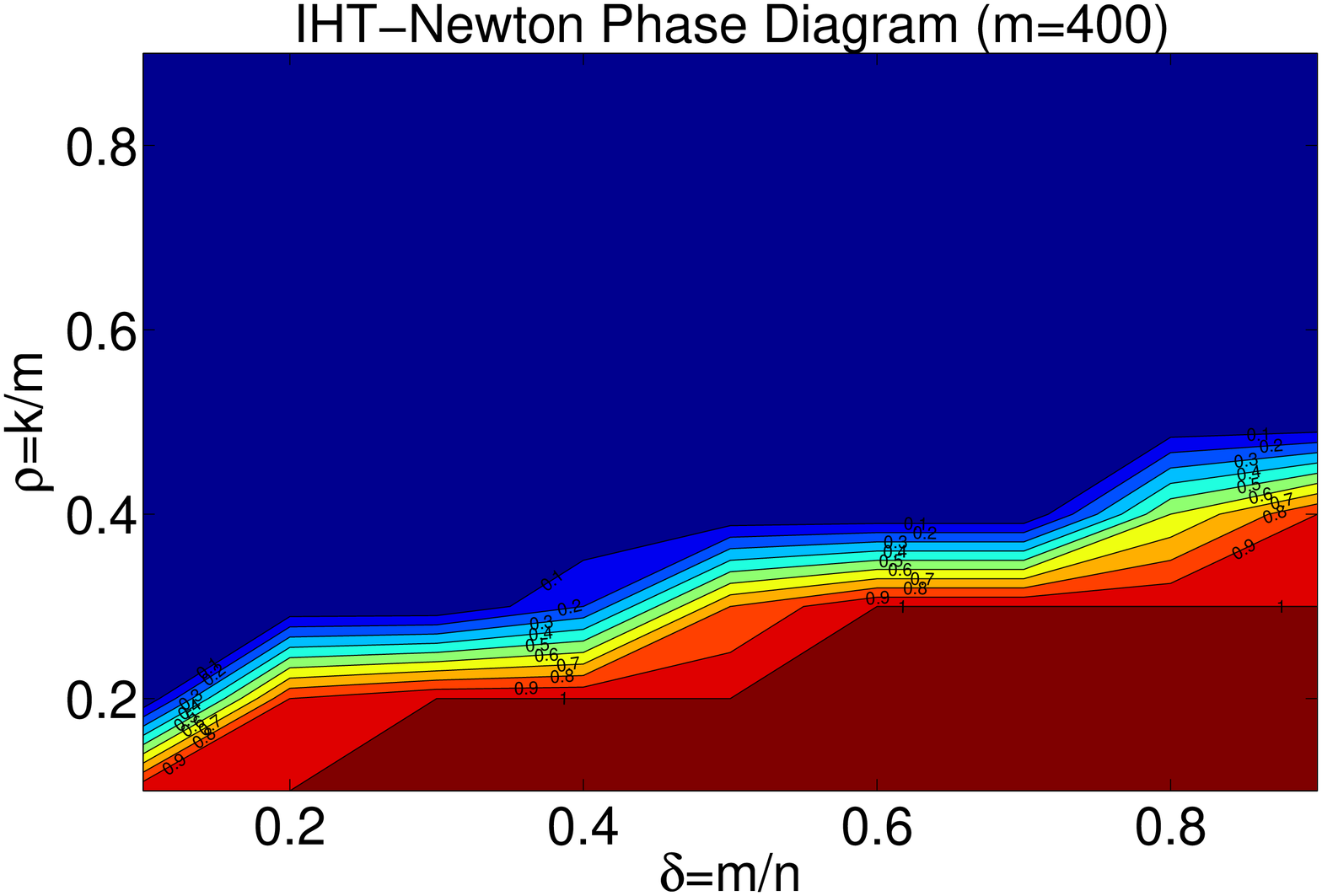}\\
(a) \ompr &(b) OMP&(c) IHT-Newton
  \end{tabular}
  \caption{Phase Transition Diagrams for different methods. Red represents high probability of success while blue represents low probability of success. Clearly, OMPR recovers correct solution for a much larger region of the plot than OMP and is comparable to IHT-Newton. (Best viewed in color)}
\label{fig:phase}
\end{figure*}
\subsection{Phase Transition Diagrams}
We first compare different methods using phase transition diagrams which are commonly used in compressed sensing literature to compare different methods~\citep{MalekiD10}. We first fix the number of measurements to be $m=400$ and generate different problem sizes by varying $\rho=k/m$ and $\delta=m/n$. For each problem size $(m,n,k)$, we generate random $m\times n$ Gaussian measurement matrices and $k$-sparse random vectors. We then estimate the probability of success of each of the method by applying the method to 100 randomly generated instances. A method is considered successful for a particular instance if it recovers the underlying $k$-sparse vector with at most $1\%$ relative error. 

In Figure~\ref{fig:phase}, we show the phase transition diagram of our \ompr method as well as that of OMP and IHT-Newton (with step size 1). The plots shows probability of successful recovery as a function of $\rho=m/n$ and $\delta=k/m$. Figure~\ref{fig:phase} (a) shows color coding of different success probabilities; red represents high probability of success while blue represents low probability of success. Note that for Gaussian measurement matrices, the RIP constant $\delta_{2k}$ is less than a fixed constant if and only if $m=C k \log (n/k)$, where $C$ is a universal constant. This implies that $\frac{1}{\delta}=C \log \rho$ and hence a method that recovers for high $\delta_{2k}$  will have a large fraction in the phase transition diagram where successful recovery probability is high. We observe this phenomenon for both \ompr and IHT-Newton method which is consistent with their respective theoretical guarantees (see Theorem~\ref{thm:family}). On the other hand, as expected, the phase transition diagram of OMP has a negligible fraction of the plot that shows high recovery probability. 
\subsection{Performance for Noisy or Under-sampled Observations}
Next, we empirically compare performance of \ompr to various existing compressed sensing methods. As shown in the phase transition diagrams in Figure~\ref{fig:phase}, \ompr provides comparable recovery to the IHT-Newton method for noiseless cases. Here, we show that \ompr is fairly robust under the noisy setting as well as in the case of under-sampled observations, where the number of observations is much smaller than what is required for exact recovery. 

\begin{figure*}[t]
\hspace*{-20pt}
\begin{tabular}{ccc}
\begin{minipage}{.33\linewidth}
    \includegraphics[width=\linewidth]{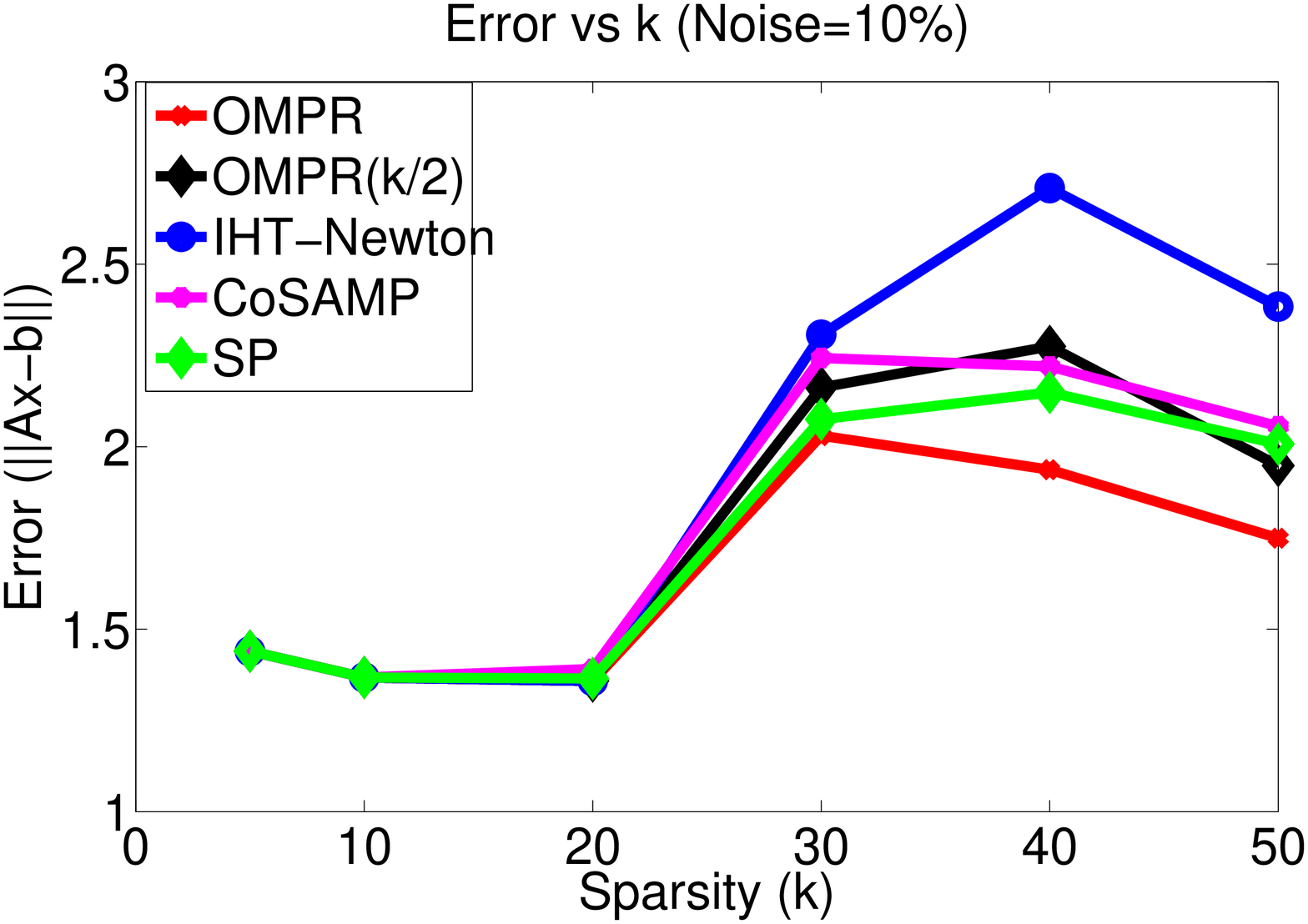}
\end{minipage}&
\hspace*{-20pt}
\begin{minipage}{.33\linewidth}
    \includegraphics[width=\linewidth]{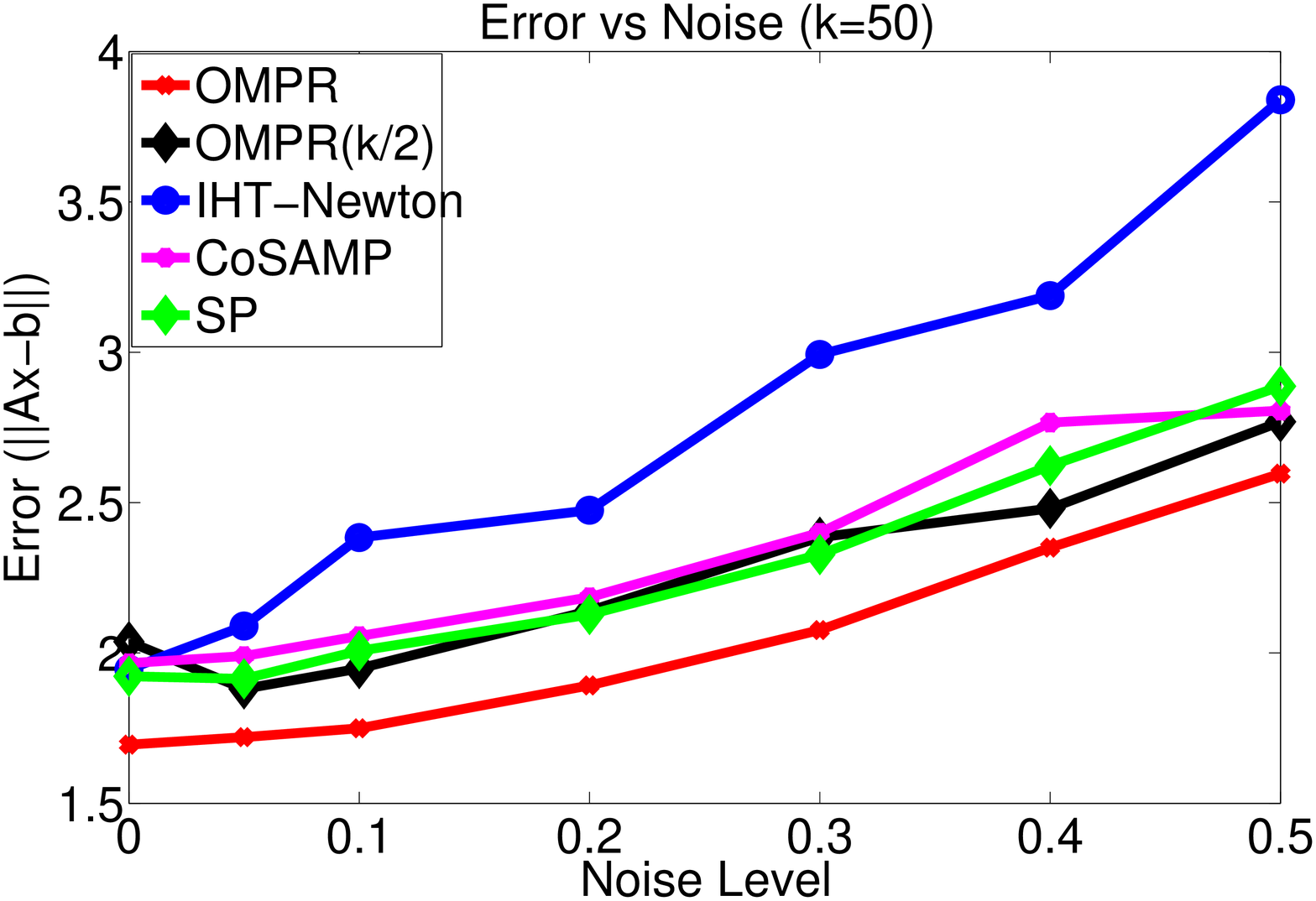}
\end{minipage}&
\begin{minipage}{.34\linewidth}
{\small
\begin{tabular}{|c|c|c|c|}\hline
Noise/ k&10&30&50\\\hline
0.00  & 0.00(0.0) & -0.21(0.6) & 0.25(0.3)\\ \hline
0.05  & 0.00(0.0) & 0.13(0.3) & 0.37(0.3)\\ \hline
0.10  & 0.00(0.0) & 0.28(0.3) & 0.63(0.4)\\ \hline
0.20  & 0.03(0.0) & 0.62(0.2) & 0.58(0.5)\\ \hline
0.30  & 0.18(0.1) & 0.92(0.3) & 0.92(0.6)\\ \hline
0.40  & 0.31(0.1) & 1.19(0.3) & 0.84(0.5)\\ \hline
0.50  & 0.37(0.1) & 1.48(0.3) & 1.24(0.6)\\ \hline
\end{tabular}
}
\end{minipage}\\
(a)&(b)&(c)
\end{tabular}
\caption{Error in recovery ($\|A\x-b\|$) of $n=3000$ dimensional vectors from $m=200$ measurements. (a): Error incurred by various methods as the sparsity level $k$ increases. Note that \ompr\ incurs the least error as it provably converges to at least a local minima for {\em fixed} step size $\eta=1$. (b): Error incurred by various methods as the noise level increases. Here again OMPR performs significantly better than the existing methods. (c): Difference in error incurred by IHT-Newton  and \ompr, i.e., Error(IHT-Newton)-Error(\ompr). Numbers in bracket denote confidence interval at 95\% significance level.}
\label{fig:cs_noise}
\end{figure*}

For this experiment, we generate random Gaussian measurement matrix of size $m=200, n=3000$. We then generate random binary vector $x$ of sparsity $k$ and add Gaussian noise to it. Figure~\ref{fig:cs_noise} (a) shows recovery error ($\|A\x-b\|$) incurred by various methods for increasing $k$ and noise level of $10\%$. Clearly, our method outperforms the existing methods, perhaps a consequence of guaranteed convergence to a local minima for {\em fixed} step size $\eta=1$. Similarly, Figure~\ref{fig:cs_noise} (b) shows recovery error incurred by various methods for fixed $k=50$ and varying noise level. Here again, our method outperforms existing methods and is more robust to noise. Finally, in Figure~\ref{fig:cs_noise} (c) we show difference in error incurred along with confidence interval (at $95\%$ signficance level) by IHT-Newton and \ompr for varying levels of noises and $k$. Our method is better than IHT-Newton (at $95\%$ signficance level) in terms of recovery error in around 30 cells of the table, and is not worse in any of the cells but one. 
\begin{figure*}[t]
  \centering
  \begin{tabular}{ccc}
    \includegraphics[width=.33\textwidth]{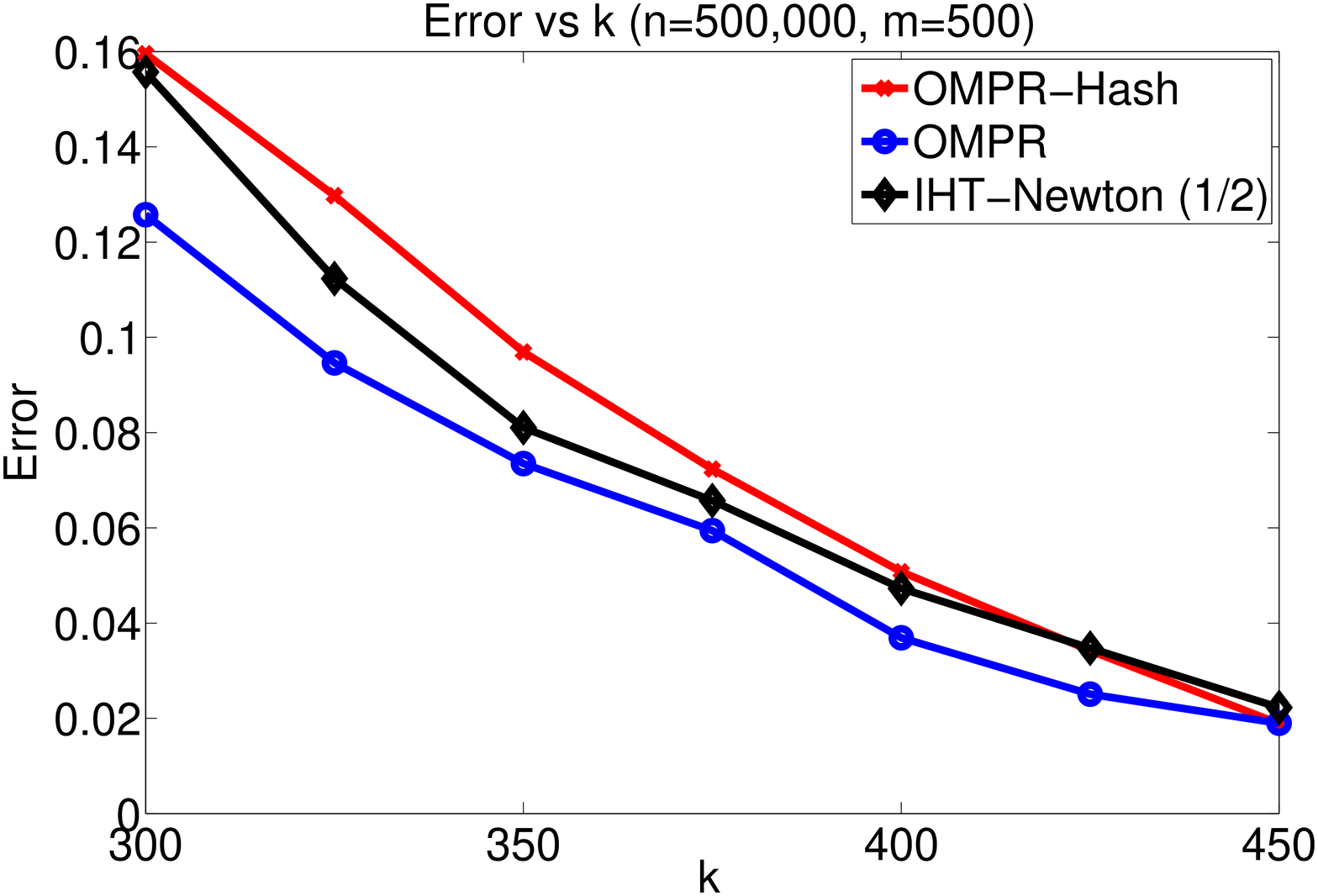}&
    \includegraphics[width=.33\textwidth]{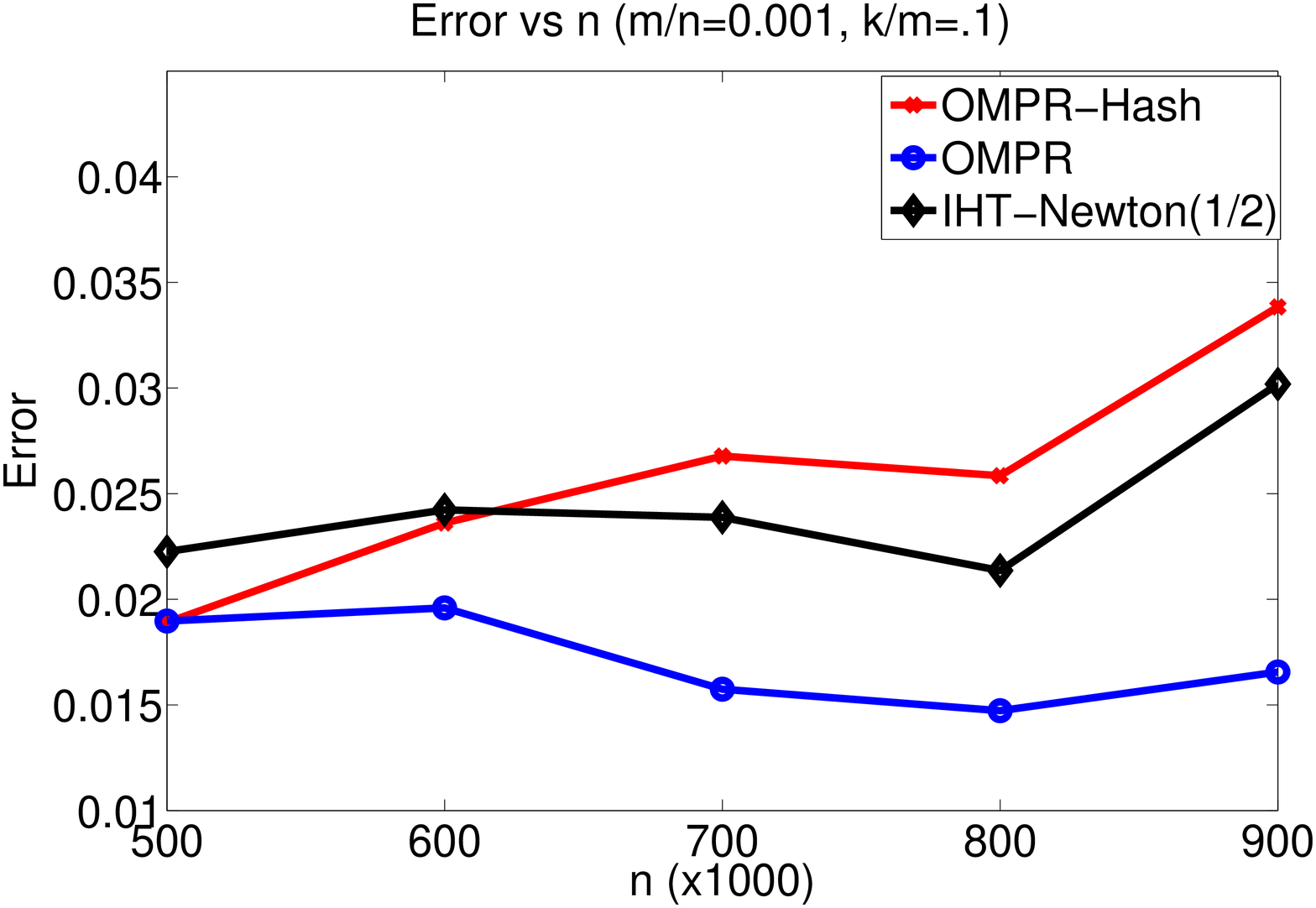}&
    \hspace*{-10pt}\includegraphics[width=.33\textwidth]{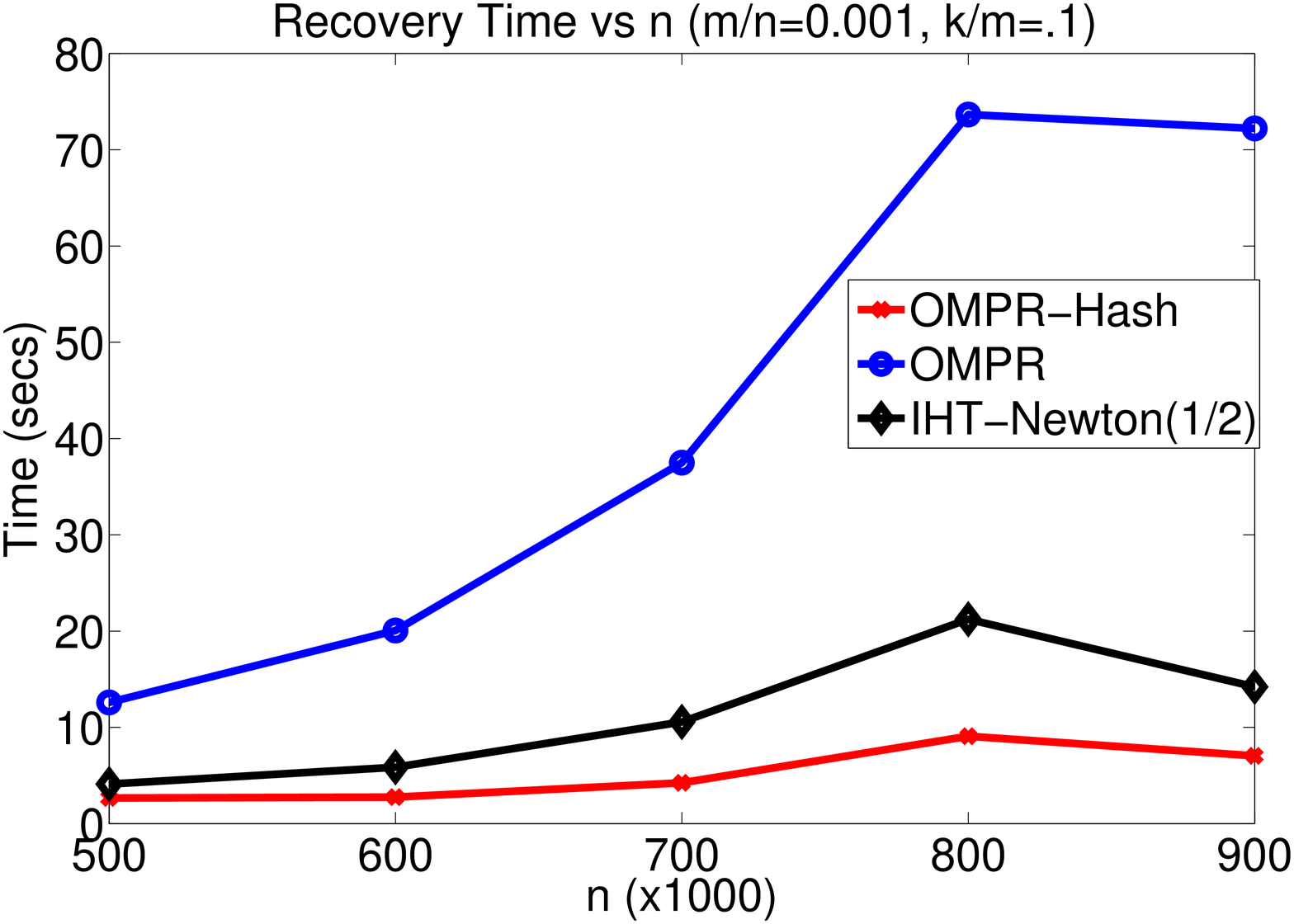}\\
(a)&(b)&(c)
  \end{tabular}
  \caption{(a): Error ($\|A\x-b\|$) incurred by various methods as $k$ increases. The measurements $\b=\A\x$ are computing by generating $\x$ with support size $m/10$. (b),(c): Error incurred and time required by various methods to recover vectors of support size $0.1m$ as $n$ increases. IHT-Newton(1/2) refers to the IHT-Newton method with step size $\eta=1/2$.}
\label{fig:lsh}
\end{figure*}
\subsection{Performance of LSH based implementation}
Next, we empirically study recovery properties of our LSH based implementation of \ompr (\lompr) in the following real-time setup: Generate a random measurement matrix from the Gaussian ensemble and construct hash tables offline using hash functions specified in Section~\ref{sec:lsh}. Next, during the reconstruction stage, measurements arrive one at a time and the goal is to recover the underlying signal accurately in real-time.
 For our experiments, we generate measurements using random sparse vectors and then report recovery error $\|\A\x-\b\|$ and computational time required by each of the method averaged over $20$ runs.  

In our first set of experiments, we empirically study the performance of different methods as $k$ increases. Here, we fix $m=500$, $n=500,000$ and generate measurements using $n$-dimensional random vectors of support set size $m/10$.  We then run different methods to estimate vectors $\x$ of support size $k$ that minimize $\|\A\x-\b\|$. For our \lompr method, we use $s=20$ bits hash-keys and generate $q=\sqrt{n}$ hash-tables. Figure~\ref{fig:lsh} (a) shows the error incurred by \ompr, \lompr, and IHT-Newton for different $k$ (recall that $k$ is an input to both \ompr and IHT-Newton). Note that although \lompr performs an approximation at each step, it is still able to achieve error similar to \ompr and IHT-Newton. Also, note that since the number of measurements are not enough for exact recovery by the IHT-Newton method, it typically diverges after a few steps. As a result, we use IHT-Newton with step size $\eta=1/2$ which is always guaranteed to monotonically converge to at least a local minima (see Theorem~\ref{thm:family}). In contrast, in \ompr and \lompr can always set step size $\eta$ aggressively to be $1$. 

Next, we evaluate \lompr as dimensionality of the data $n$ increases. For \lompr, we use $s=\log_2(n)$ hash-keys and $q=\sqrt{n}$ hash-tables. Figures~\ref{fig:lsh}(b) and (c) compare error incurred and time required by \lompr with \ompr and IHT-Newton. Here again we use step size $\eta=1/2$ for IHT-Newton as it does not converge for $\eta=1$. Note that \lompr is an order of magnitude faster than \ompr while incurring slightly higher error.  \lompr is also nearly $2$ times faster than IHT-Newton. 



\newpage
\bibliographystyle{plain}
\bibliography{cs_paper}

\clearpage
\appendix
\section{Proofs related to OMPR: Exact Recovery Case}
\label{app:ompr}

Let us denote the objective function by $f(x)=\frac{1}{2}\|\A\x-\b\|^2$.
Let $\curr$ denote the support set of $\curriter$ and $\true$ be the support set of $\truevec$. Define the sets
\begin{align*}
\fa &= \curr \backslash \true &&\text{(false alarms)}\\
\md &= \true \backslash \curr &&\text{(missed detections)}\\
\co &= \curr \cap \true &&\text{(correct detections)}\ .
\end{align*}
As the algorithms proceed, elements get in and move in and out of the current set $\curr$. Let us give names to the
set of found and lost elements as we move from $\curr$ to $\next$:
\begin{align*}
\found &= \next \backslash \curr &&\text{(found)} \\
\lost &= \curr \backslash \next &&\text{(lost)}\ .
\end{align*}

We first state two technical lemmas that we will need. These can be found in \cite{NeedellT09}.
 
\begin{lemma}
For any $S \subset [n]$, we have,
$$\|I-\A_S^T\A_S\| \leq \delta_{|S|}.$$
\label{lem:lem1}
\end{lemma}

\begin{lemma}
For any $S,T \subset [n]$ such that $S\cap T=\emptyset$, we have,
$$\|\A_S^T\A_T\|_2\leq \delta_{|S\cap T|}.$$
\label{lem:lem2}
\end{lemma}

\subsection*{Proof of Theorem 3}

\begin{lemma}
\label{claim:1}
Let $\delta_{2k}<1-\frac{1}{2\eta}$. 
Then, in \omprf($l$), $$0<2(2\eta-\frac{1}{1-\delta_{2k}})f(\x^t)\leq \|\zmd\|^2-\|\xfa\|^2.$$
\end{lemma}
\begin{proof}
Since $\xcurr$ is the solution to the least squares problem $\min_{\x} \|\Acurr \x-\b\|^2$, 
 \begin{equation}
 \Acurr^T(\Acurr\xcurr-\b)=\mathbf{0}.
 \label{eq:ls}
 \end{equation}

Now, note that 
\begin{align}
f(\x^t)&=\frac{1}{2}\|\Acurr\xcurr-\Atrue\xtrue\|^2,\nonumber\\
&=\frac{1}{2}((\xcurr)^T\Acurr^T(\Acurr\x^t-\Atrue\x^*)-(\xtrue)^T\Atrue^T(\Acurr\x^t-\Atrue\x^*)),\nonumber\\
&=-\frac{1}{2}(\xmd)^T\Amd^T(\Acurr\xcurr-\Atrue\xtrue),\qquad \text{by \eqref{eq:ls}}\nonumber\\
&=\frac{1}{2\eta}(\xmd)^T\zmd.\qquad \text{by \eqref{eq:zt}}
\label{eq:znio}
\end{align}

Hence, 
\begin{align}
\|\xmd-\zmd\|^2 &=\|\xmd\|^2+\|\zmd\|^2-2(\xmd)^T\zmd \nonumber \\
&=\|\xmd\|^2+\|\zmd\|^2-4\eta f(\x^t). 
\label{eq:xznio}
\end{align}
That is,
\begin{align*}
4\eta f(\x^t)&\leq \|\xmd\|^2+\|\zmd\|^2,\\
&\leq\|\xmd\|^2+\|\xfa\|^2+\|\xcurrco-\xtrueco\|^2-\|\xfa\|^2+\|\zmd\|^2,\\
&=\|\curriter -\truevec\|^2+\|\zmd\|^2-\|\xfa\|^2,\\
&\leq \frac{1}{1-\delta_{2k}}\|\A(\curriter - \truevec)\|^2+\|\zmd\|^2-\|\xfa\|^2,\qquad \text{by RIP}\\
&=\frac{2}{1-\delta_{2k}}f(\x^t)+\|\zmd\|^2-\|\xfa\|^2, 
\end{align*}
where the third line follows from the fact that $\md$, $\fa$, and $\co$ are disjoint sets. 

As $f(\x^t)> 0$ and $\delta_{2k}<1-\frac{1}{2\eta}$, the above inequality implies
$$0<2(2\eta-\frac{1}{1-\delta_{2k}})f(\x^t)\leq \|\zmd\|^2-\|\xfa\|^2.$$
\end{proof}

Next, we provide a lemma that bounds the function value $f(\x^t)$ in terms of missed detections $\md$ and also $\zmd$. 
\begin{lemma}
\label{lem:flb}
Let $f(\x^t)=\frac{1}{2}\|A\x^t-b\|^2$, $b=A\x^*$, $\delta_{2k}<1-\frac{1}{2\eta}$ and $\eta<1$. Then, at each step,
\begin{equation}
\label{eq:flblemma}
\frac{(1-\eta)^2}{\eta}\| \xmd \|^2\ \leq\ f(\x^t)\ \leq\ \frac{1}{4\eta(1-\eta)^2}\|\zmd\|^2
\end{equation}
\end{lemma}
\begin{proof}
  Now, using Lemma 2 of \citep{DaiM09} with $I = \md$, $J = \curr$, $y = \A_{\md}\truevec_{\md}$ we get
\begin{align}
\notag
f(\x^t) &= \tfrac{1}{2}\|\A\curriter - \b\|^2 \\
&=\tfrac{1}{2}\|\A_{\curr}(\curriter-\truevec)_{\curr} - \A_{\md}\truevec_{\md}\|^2 \\
\notag
&\ge \tfrac{1}{2} \left( 1 - \frac{\delta_{2k}}{1-\delta_k} \right)^2 \|\A_{\md}\truevec_{\md}\|^2 \\
\notag
&\ge \tfrac{1}{2} \left( 1 - \frac{\delta_{2k}}{1-\delta_k} \right)^2 (1-\delta_k) \| \xmd \|^2 \qquad \text{by RIP}\\
\notag
&\ge \tfrac{1}{2} \left( 1 - \frac{\delta_{2k}}{1-\delta_{2k}} \right)^2 (1-\delta_{2k}) \| \xmd \|^2 \\
&\ge \frac{(1-2\delta_{2k})^2}{2(1-\delta_{2k})}  \| \xmd \|^2 
\label{eq:flb1}
\end{align}
The assumption that $\delta_{2k}<1-\tfrac{1}{2\eta}$ and $\eta < 1$ implies that $\delta_{2k} < 1-\tfrac{1}{2\eta} < 1/2$. The function
$\alpha \mapsto (1-2\alpha)^2/(2(1-\alpha))$ is decreasing on $[0,1/2]$ and hence~\eqref{eq:flb1} implies
\begin{align}
f(\x^t) \ge \frac{\left( 1-2(1-\frac{1}{2\eta}) \right)^2}{ 2(1 - 1 + \frac{1}{2\eta}) }\| \xmd \|^2 = \frac{(1-\eta)^2}{\eta}\| \xmd \|^2.
\label{eq:flb2}
\end{align}

Next, using \eqref{eq:znio} and Cauchy-Schwarz inequality:
\begin{equation}
  \|\zmd\|^2\geq 4\eta^2 \frac{f(\x^t)^2}{\|\xmd\|^2}.
\end{equation}
The result now follows using the above equation with \eqref{eq:flb2}.
\end{proof}

\begin{lemma}
\label{claim:2}
Let $\delta_{2k}<1-\frac{1}{2\eta}$ and $1/2<\eta<1$. 
Then assuming $f(x^t)>0$, at least one new element is found i.e. $\found \neq \emptyset$.
Furthermore, $\|\yfound\|^2>\frac{l}{k}cf(\x^t)$, where $c=\min(4\eta(1-\eta)^2,2(2\eta-\frac{1}{1-\delta_{2k}}))>0$ is a constant.
\end{lemma}
\begin{proof}
We consider the following three exhaustive cases:
\begin{enumerate}
\item  $|F_t|<l$ and $|F_t|<|MD_t|$: Let $S\subseteq |\md\backslash \found|$, s.t., $|S|=|\found|-|\md\cap \found|$. Now, $$|S\cup (\md\cap \found)|=|\found|,\ \ \ |(\md\backslash \found)\backslash S|=|\md|-|\found|.$$
Now, as $\y_{\found}$ consists of top $\found$ elements of $\zmd$:
\begin{equation}
\label{eq:smdf}
\|\z^{t+1}_{S\cup (\md\cap \found)}\|^2\leq \|\y_{\found}\|^2.
\end{equation}
Furthermore, since $|F_t|<l$, hence every element of $\z^{t+1}_{\md\backslash \found}$ is smaller in magnitude than every element of $\x^t_{\fa\backslash \lost}$, otherwise that element should have been included in $\found$. Furthermore, $|\md|-|\found|=|\fa|-|\lost|\leq |\fa\backslash \lost|$. Hence, 
\begin{equation}
\label{eq:mdfs}
\|\z^{t+1}_{(\md\backslash \found)\backslash S}\|^2\leq \|\x^{t}_{\fa\backslash \lost}\|^2\leq  \|\x^{t}_{\fa}\|^2,
\end{equation}
Adding \eqref{eq:smdf} and \eqref{eq:mdfs}, we get: 
\begin{equation}
\label{eq:zyfa}
  \|\z^{t+1}_{\md}\|^2\leq \|\yfound\|^2+ \|\x^{t}_{\fa}\|^2.
\end{equation}
Using above equation along with Lemma~\ref{claim:1}, we get:
\begin{equation}
  \label{eq:c1}
  \|\yfound\|^2\geq 2\left(2\eta-\frac{1}{1-\delta_{2k}}\right)f(\x^t).
\end{equation}
Now, note that if $|\found|=0$, then $\yfound=0$ implying that $f(\x^t)=0$. Hence, at least one new element is added, i.e., $\yfound\neq \emptyset$. 
\item $|F_t|=l< |MD_t|$: By definition of $\yfound$:
$$\frac{\|\yfound\|^2}{|F_t|}\geq \frac{\|\zmd\|^2}{|\md|}.$$
Hence, using Lemma~\ref{lem:flb} and the fact that $|\found|=l$:
\begin{equation}
\label{eq:c2}
\|\yfound\|^2\geq \frac{l}{|\md|}4\eta(1-\eta)^2f(\x^t)\geq \frac{l}{k}4\eta(1-\eta)^2f(\x^t),
\end{equation}
as $|\md|\leq k$. 
\item $|\found|\geq |\md|$: Since, $\yfound$ is the top most elements of $\z^{t+1}$. Hence, assuming $|F_t|\geq |MD_t|$,
$$\|\yfound\|^2\geq \|\zmd\|^2.$$
Now, using Lemma~\ref{lem:flb}:
\begin{equation}
\label{eq:c3}
\|\yfound\|^2\geq 4\eta(1-\eta)^2f(\x^t).
\end{equation}
\end{enumerate}
We get the lemma by combining bounds for all the three cases, i.e., \eqref{eq:c1}, \eqref{eq:c2}, \eqref{eq:c3}.
\end{proof}

Now we give a complete proof of Theorem~\ref{thm:family}.
\begin{proof}
We have, 
\begin{align}
f(\y^{t+1})-f(\curriter)&=(\y^{t+1}-\curriter)^T\A^T\A(\curriter-\truevec)+1/2\|\A(\y^{t+1}-\curriter)\|^2,\nonumber\\
&\leq (\y^{t+1}-\curriter)^T\A^T\A(\curriter-\truevec)+\frac{(1+\delta_{2l})}{2} (\|\yfound\|^2+\|\xlost\|^2).
\label{eq:fdiff}
\end{align}
where the second inequality follows by using the fact that $\y^{t+1}_{\next \cap \curr}=\curriter_{\next\cap \curr}$ and using RIP of order $2l$
(since $|\supp(\y^{t+1}-\curriter)| = |\found \cup \lost| \le 2l$). 

Since $\x^t_{I_t}$ is obtained using least squares, $$\A^T_{\curr}\A(\curriter-\truevec)=\zero.$$
Thus, $\A^T_{\lost}\A(\curriter-\truevec)=\zero$, because $\lost \subseteq \curr$.
Next, note that $$\yfound = -\eta \Afound^T\A(\curriter-\truevec).$$
Hence, 
\begin{align}
f(\y^{t+1})-f(\curriter)&\leq \left(\frac{1+\delta_{2l}}{2}-\frac{1}{\eta}\right)\|\yfound\|^2+ \frac{1+\delta_{2l}}{2} \|\xlost\|^2.
\label{eq:fdiff1}
\end{align}

Furthermore,  since $\y^{t+1}$ is chosen based on the $k$ largest entries in $\z^{t+1}_{\J_{t+1}}$, we have,
$$\|\yfound\|^2 = \|\z^{t+1}_{\found}\|^2 \geq \|\z^{t+1}_{\lost}\|^2 = \|\xlost\|^2\ .$$

Plugging this into \eqref{eq:fdiff1}, we get:
\begin{align*}
f(\y^{t+1})-f(\curriter)\leq \left(1+\delta_{2l} - \frac{1}{\eta} \right)\|\yfound\|^2 \ . 
\end{align*}

Now, using Lemma~\ref{claim:2}, $\|\yfound\|^2\geq \frac{l}{k}cf(\x^t)$ and therefore,
\begin{align*}
f(\nextiter)-f(\curriter) \leq f(\y^{t+1})-f(\x^t) \leq -\alpha\frac{l}{k}f(\x^t)
\end{align*}
where $\alpha=c\left(1+\delta_{2l} - \frac{1}{\eta} \right)>0$ since $\eta(1+\delta_{2l}) < 1$. Hence,
$$f(\nextiter)\leq (1-\alpha\frac{l}{k})f(\curriter) \le e^{-\alpha\frac{l}{k}} f(\curriter).$$
The above inequality shows that at each iteration \omprf($l$) reduces the objective function value by a fixed multiplicative factor.
Furthermore, if $\x^0$ is chosen to have entries bounded by $1$, then $f(\x^0) \leq (1+\delta_{2k})k$.
Hence, after $O(\frac{k}{l}\log ((1+\delta_{2k}) k/\epsilon))$ iterations, the function value reduces to $\epsilon$, i.e., $f(\x^t)\leq \epsilon$. 
\end{proof}

\section{Proofs related to the LSH Section}
\label{app:lsh}
\begin{lemma}
  \label{lemma:nnsn}
Let $\|\x\|=1$ for all points $\x$ in our database. Let $x^*$ be the nearest neighbor to $\r$ in $L_2$ distance metric, and let $\r^T\x^*\geq c>0$. Then, a $(1+\alpha\epsilon)$-nearest neighbor to $\r$ is also a $(1-\epsilon)$-similar neighbor to $\r$, where $\alpha\leq \frac{2c}{1+\r^T\r-2c}$. 
\end{lemma}
\begin{proof}
  Let $x'$ be a $(1+\alpha\epsilon)$-nearest neighbor to $\r$, then:
$$\|\x'-\r\|^2\leq (1+\alpha\epsilon)\|\x^*-\r\|^2.$$
Using $\|\x'\|=\|\x^*\|=1$ and simplifying, we get:
\begin{align*}
\r^T\x'&\geq (1-\epsilon)\r^T\x^*+(\alpha+1)\epsilon\r^T\x^*-\frac{\alpha\epsilon}{2}(1+\r^T\r),\\
&\geq (1-\epsilon)\r^T\x^*+((\alpha+1)c-\frac{\alpha}{2}(1+\r^T\r))\epsilon. 
\end{align*}
Hence, $\x'$ is a $(1-\epsilon)$-approximate similar neighbor to $\r$ if:
$$(\alpha+1)c\geq \frac{\alpha}{2}(1+\r^T\r).$$
The result follows after simplification. 
\end{proof}
We now provide a proof of Theorem 7. 
\begin{proof}
Let us first consider a single step of \ompr. Now, similar to Lemma~\ref{claim:1}, we can show that if $\delta_{2k}<1/4 - \gamma$ and $\eta=1-\gamma$, $\gamma>0$, then $\|\zmd\|^2> \frac{3}{2}\|\xfa\|^2$. Setting $\epsilon=1-\sqrt{\frac{2}{3}}$, implies that $(1-\epsilon) \max |\zmd|\geq \min |\xfa|$, i.e., a $(1-\epsilon)$-similar neighbor to $\max |\zmd|$ will still lead to a constant decrease in the objective function. 

So, the goal is to ensure that with probability $1-\delta$, $\delta>0$, for all the $O(k)$ iterations, our LSH method returns at least a $(1-\epsilon)$-similar neighbor to $\max |\zmd|$ where $\epsilon=1-\sqrt{\frac{2}{3}}$. To this end, we need to ensure that at each step $t$, LSH finds at least a $(1-\epsilon)$-similar neighbor to $\max |\zmd|$ with probability at least $1-\delta/k$. Using Lemma~\ref{lemma:nnsn}, we need to find a $(1+\alpha\epsilon)$-nearest neighbor to $\max |\zmd|$, where 
$$\alpha\leq \frac{2c}{1+\r^T\r-2c},$$
and $\r^T\x^*\geq c$. Using Lemma~\ref{claim:2}, $\alpha=O(1/k)$. Hence the result now follows using Theorem 6 (main text). 
\end{proof}

\section{Extension to Noisy Case}
\label{app:noisy}
In this section, we consider the noisy case in which our objective function is $f(x)=\tfrac{1}{2} \|\A\x-\b\|^2$, where $b=\A\x^*+e$ and $e\in \mathbb{R}^m$ is the ``noise'' vector. 

Let $\curr$ denote the support set of $\curriter$ and $\true$ be the support set of $\truevec$. Define the sets
\begin{align*}
\fa &= \curr \backslash \true &&\text{(false alarms)}\\
\md &= \true \backslash \curr &&\text{(missed detections)}\\
\co &= \curr \cap \true &&\text{(correct detections)}\ .
\end{align*}

\begin{lemma}
\label{claim:1_noise}
Let $f(\x^t)\geq \frac{C}{2}\|e\|^2$ and $\delta_{2k}<1-\frac{1}{2D\eta}$, where $D=\frac{C-\sqrt{C}}{(\sqrt{C}+1)^2}$.
Then, $$\|\zmd\|^2- \|\xfa\|^2\geq cf(\x^t),$$ where $c=2\frac{(\sqrt{C}+1)^2}{C}(2\eta D-\frac{1}{1-\delta_{2k}})>0$. 
\end{lemma}
\begin{proof}
Since $\xcurr$ is the solution to the least squares problem $\min_{\x} \|\Acurr \x-\b\|^2$, 
 \begin{equation}
 \Acurr^T(\Acurr\xcurr-\b)=0.
 \label{eq:ls_n}
 \end{equation}

Now, note that 
\begin{align}
f(\x^t)&=\frac{1}{2}\|\Acurr\xcurr-b\|^2,\nonumber\\
&=\frac{1}{2}((\xcurr)^T\Acurr^T(\Acurr\x^t-b)-b^T(\Acurr\x^t-b)),\nonumber\\
&=-\frac{1}{2}b^T(\Acurr\xcurr-b),\nonumber\\
&=-\frac{1}{2}(\x^*_{\md})^T\A_{\md}^T(\Acurr\xcurr-b)-\frac{1}{2}e^T(\Acurr\xcurr-b),\nonumber\\
&=\frac{1}{2\eta}(\xmd)^T\zmd-\frac{1}{2}e^T(\Acurr\xcurr-b),
\label{eq:znio_n}
\end{align}
where the third equality follows from \eqref{eq:ls_n}. 

Now, 
\begin{align}
\|\xmd-\zmd\|^2 &=\|\xmd\|^2+\|\zmd\|^2-2(\xmd)^T\zmd\nonumber \\
&=\|\xmd\|^2+\|\zmd\|^2-4\eta (f(\x^t)+\frac{1}{2}e^T(\Acurr\xcurr-b))
\label{eq:xznio_noise1}
\end{align}
So,
\begin{align*}
0&\leq \|\xmd\|^2+\|\xfa\|^2-\|\xfa\|^2+\|\zmd\|^2-4\eta (f(\x^t)+\frac{1}{2}e^T(\Acurr\xcurr-b)),\\
&\leq \|\xmd\|^2+\|\xfa\|^2+\|\xcurrco-\xtrueco\|^2-\|\xfa\|^2+\|\zmd\|^2 -4\eta(f(\x^t)+\frac{1}{2}e^T(\Acurr\xcurr-b)),\\
&\leq \|x^t-x^*\|^2-\|\xfa\|^2+\|\zmd\|^2-4\eta(f(\x^t)+\frac{1}{2}e^T(\Acurr\xcurr-b)),\\
&\leq \frac{1}{1-\delta_{2k}}\|\A(x^t-x^*)\|^2-\|\xfa\|^2+\|\zmd\|^2-4\eta(f(\x^t)+\frac{1}{2}e^T(\Acurr\xcurr-b)),\\
&= \frac{1}{1-\delta_{2k}}\|\A(x^t-x^*)\|^2-\|\xfa\|^2+\|\zmd\|^2-4\eta(1-\frac{1}{\sqrt{C}})f(\x^t).
\end{align*}
Now, by assumption: $f(\x^t)\geq \frac{C}{2}\|e\|^2$. Hence, 
\begin{align*}
  \|\A(x^t-x^*)\|&\leq \|\A(x^t-\x^*)-e\|+\|e\|,\\
\|\A(x^t-x^*)\|^2&\leq 2(1+\frac{1}{\sqrt{C}})^2f(\x^t).
\end{align*}
Hence, 
\begin{align*}
2\left(2\eta(1-\frac{1}{\sqrt{C}})-\frac{1}{1-\delta_{2k}}(1+\frac{1}{\sqrt{C}})^2\right)f(\x^t)\leq \|\zmd\|^2-\|\xfa\|^2
\end{align*}
Now, by assumption $\delta_{2k}< 1- \frac{1}{2D\eta}$, where $D=\frac{(\sqrt{C}+1)^2}{C-\sqrt{C}}$. Hence, $c=2\frac{(\sqrt{C}+1)^2}{C}(2\eta D-\frac{1}{1-\delta_{2k}})>0$. 
\end{proof}

Next, we provide a lemma that bounds the function value $f(\x^t)$ in terms of missed detection $\md$ and also $\zmd$. 
\begin{lemma}
\label{lem:flb_n}
Let $f(\x^t)=\frac{1}{2}\|A\x^t-b\|^2\geq \frac{C}{2}\|e\|^2$, $b=A\x^*+e$, $\delta_{2k}<1-\frac{1}{2D\eta}$ and $D=\frac{C-\sqrt{C}}{(\sqrt{C}+1)^2}$. Then, at each step,
\begin{equation}
\label{eq:flblemma_n}
\frac{(1-\eta)^2C}{\eta(\sqrt{C}+1)^2}\| \xmd \|^2\ \leq\ f(\x^t)\ \leq\ \frac{1}{4\eta(1-\eta)^2}\frac{(\sqrt{C}+1)^2}{(\sqrt{C}-1)^2}\|\zmd\|^2
\end{equation}
\end{lemma}
\begin{proof}
First we lower bound $f(x^t)$:
\begin{align*}
  \sqrt{f(x^t)}&=\frac{1}{\sqrt{2}}\|Ax^t-Ax^*-e\|,\\
&\geq \frac{1}{\sqrt{2}}\left(\|Ax^t-Ax^*\|-\|e\|\right),\\
&\geq \frac{1}{\sqrt{2}}\left(\min_{x\::\: x_{\currc}=0} \|Ax-Ax^*\|-\|e\|\right),\\
&\geq \frac{1}{\sqrt{2}}\left(\frac{(1-2\delta_{2k})}{\sqrt{(1-\delta_{2k})}}  \| \xmd \|-\|e\| \right),
\end{align*}
where last equality follows from Lemma~\ref{lem:flb}. Using the above inequality with $f(x^t)\geq \frac{C}{2}\|e\|^2$, we get:
\begin{equation}
f(x^t)\geq \frac{(1-2\delta_{2k})^2C}{2(1-\delta_{2k})(\sqrt{C}+1)^2}\|\xmd\|^2. 
\label{eq:xlb3_n}
\end{equation}
The assumption that $\delta_{2k}<1-\tfrac{1}{2D\eta}$ and $D\eta < 1$ implies that $\delta_{2k} < 1-\tfrac{1}{2D\eta} < 1/2$. The function
$\alpha \mapsto (1-2\alpha)^2/(2(1-\alpha))$ is decreasing on $[0,1/2]$ and hence the above equation implies
\begin{align}
f(\x^t) \ge \frac{(1-D\eta)^2}{D\eta}\| \xmd \|^2.
\label{eq:flb2_n}
\end{align}
Now, we upper bound $f(\x^t)$. Using definition of $f(\x^t)$:
$$\frac{1}{2\eta}(\xmd)^T\zmd=f(x^t)+\frac{1}{2}e^T(\Acurr\xcurr-b).$$
Now, using Cauchy-Schwarz and $f(\curriter) \ge \frac{C}{2}\|e\|^2$,
$$\left|e^T(\Acurr\xcurr-b)\right|\leq \|e\|\|\Acurr\xcurr-b\|\leq \frac{2}{\sqrt{C}}f(x^t).$$
Hence, $$\frac{1}{2\eta}\|\xmd\|\|\zmd\|\geq \frac{1}{2\eta}(\xmd)^T\zmd\geq (1-\frac{1}{\sqrt{C}})f(x^t).$$
That is,
\begin{equation}
\|\zmd\|^2 \geq 4\eta^2\left(1-\frac{1}{\sqrt{C}}\right)^2\frac{f(x^t)^2}{\|\xmd\|^2}\geq 4\eta(1-D\eta)^2\frac{(\sqrt{C}-1)^2}{CD}f(\x^t),
\label{eq:zlb2_n}
\end{equation}
where the second inequality follows from \eqref{eq:flb2_n}. 
\end{proof}

Next, we present the following lemma that shows ``enough'' progress at each step:
\begin{lemma}
\label{claim:2_noise}
Let $f(\x^t)\geq \frac{C}{2}\|e\|^2$, $\eta<1$ and $\delta_{2k}<1-\frac{1}{2D\eta}$, where $D=1-\frac{1}{\sqrt{C}-1}$.  
Then at least one new element is found i.e. $\found \neq \emptyset$. Furthermore, $\|\yfound\|>\frac{l}{k}\alpha f(\x^t)$, where $\alpha=\min(4\eta(1-D\eta)^2\frac{(\sqrt{C}-1)^2}{CD},2\frac{(\sqrt{C}+1)^2}{C}(2\eta D-\frac{1}{1-\delta_{2k}}))>0$ is a constant.
\end{lemma}
\begin{proof}
As for the exact case, we analyse the following three exhaustive cases:
\begin{enumerate}
\item  $|F_t|<l$ and $|F_t|<|MD_t|$: Here we use the exactly similar argument as the exact case to obtain the following inequality (see \eqref{eq:zyfa}):
\begin{equation}
\label{eq:zyfa_n}
  \|\z^{t+1}_{\md}\|^2\leq \|\yfound\|^2+ \|\x^{t}_{\fa}\|^2.
\end{equation}
Using Lemma~\ref{claim:1_noise}, we get:
\begin{equation}
  \label{eq:c1_n}
  \|\yfound\|^2\geq c f(\x^t),
\end{equation}
where $c$ is as defined in Lemma~\ref{claim:1_noise}.
Now, note that if $|\found|=0$, then $\yfound=0$ implying that $f(\x^t)=0$. Hence, at least one new element is added, i.e., $\yfound\neq \emptyset$. 
\item $|F_t|=l< |MD_t|$: By definition of $\yfound$:
$$\frac{\|\yfound\|^2}{|F_t|}\geq \frac{\|\zmd\|^2}{|\md|}.$$
Hence, using Lemma~\ref{lem:flb_n} and the fact that $|\found|=l$:
\begin{equation}
\label{eq:c2_n}
\|\yfound\|^2\geq \frac{l}{|\md|}4\eta(1-D\eta)^2\frac{(\sqrt{C}-1)^2}{CD}f(\x^t)\geq \frac{l}{k}4\eta(1-D\eta)^2\frac{(\sqrt{C}-1)^2}{CD}f(\x^t),
\end{equation}
as $|\md|\leq k$. 
\item $|\found|\geq |\md|$: Since, $\yfound$ is the top most elements of $\z^{t+1}$. Hence, assuming $|F_t|\geq |MD_t|$,
$$\|\yfound\|^2\geq \|\zmd\|^2.$$
Now, using Lemma~\ref{lem:flb_n}:
\begin{equation}
\label{eq:c3_n}
\|\yfound\|^2\geq 4\eta(1-D\eta)^2\frac{(\sqrt{C}-1)^2}{CD}f(\x^t).
\end{equation}
\end{enumerate}
We get the lemma by combining bounds for all the three cases, i.e., \eqref{eq:c1_n}, \eqref{eq:c2_n}, \eqref{eq:c3_n}.
\end{proof}

Now, we provide a proof of Theorem~\ref{thm:ompr_noisy}. 
\begin{proof}
We have, 
\begin{align}
f(\y^{t+1})-f(\curriter)&=(\y^{t+1}-\curriter)^T\A^T(\A\curriter-b)+1/2\|\A(\y^{t+1}-\curriter)\|^2,\nonumber\\
&\leq (\y^{t+1}-\curriter)^T\A^T(\A\curriter-b)+\frac{(1+\delta_{2l})}{2} (\|\yfound\|^2+\|\xlost\|^2).
\label{eq:fdiff_n}
\end{align}
where the second inequality follows by using the fact that $\y^{t+1}_{\next \cap \curr}=\curriter_{\next\cap \curr}$ and using RIP of order $2l$
(since $|\supp(\y^{t+1}-\curriter)| = |\found \cup \lost| \le 2l$). 

Since $\x^t_{I_t}$ is obtained using least squares, $$\A^T_{\curr}(\A\curriter-b)=\zero.$$
That is, $\A^T_{\lost}(\A\curriter-b)=\zero$, because $\lost \subseteq \curr$.
Next, note that $$\yfound = -\eta \Afound^T(\A\curriter-b).$$
Hence, 
\begin{align}
f(\y^{t+1})-f(\curriter)&\leq \left(\frac{1+\delta_{2l}}{2}-\frac{1}{\eta}\right)\|\yfound\|^2+ \frac{1+\delta_{2l}}{2} \|\xlost\|^2.
\label{eq:fdiff1_n}
\end{align}

\noindent Furthermore,  since $\y^{t+1}$ is chosen based on largest entries in $\z^{t+1}_{\J_{t+1}}$, we have,
$$\|\yfound\|^2 = \|\z^{t+1}_{\found}\|^2 \geq \|\z^{t+1}_{\lost}\|^2 = \|\xlost\|^2\ .$$

\noindent Plugging this into \eqref{eq:fdiff1_n}, we get:
\begin{align*}
f(\y^{t+1})-f(\curriter)\leq \left(1+\delta_{2l} - \frac{1}{\eta} \right)\|\yfound\|^2 \ . 
\end{align*}

\noindent Now, using Lemma~\ref{claim:2_noise}, $\|\yfound\|^2\geq \alpha f(\x^t) > 0$ and therefore,
\begin{align*}
f(\nextiter)-f(\curriter) &\leq f(\y^{t+1})-f(\x^t) \\
&\leq -c' \frac{l}{k}f(\x^t) \ ,
\end{align*}
where $c'=\frac{1 - \eta(1+\delta_{2l})}{\eta(1+\delta_{2l})} \, \alpha>0$ since $\eta(1+\delta_{2l}) < 1$. The above inequality shows that at each iteration \omprf($l$) reduces the objective function value by a fixed multiplicative factor. Furthermore, if $\x^0$ is chosen to have entries bounded by $1$, then $f(\x^0) \leq O((1+\delta_{2k})k+ \|e\|^2)$.
Hence, after $O(\frac{k}{l}\log ((k+\|e\|^2)/\epsilon))$ iterations, the function value reduces to $C\|e\|^2/2+\epsilon$. 
\end{proof}

\newcommand{\ts}{Two-stage}
\section{Analysis of 2-stage Algorithms}
\label{sec:2stage}
In this section, we consider the family of two-stage hard thresholding algorithms (see Algorithm~\ref{alg:twostage}) introduced by \cite{MalekiD10}. 
\begin{algorithm}[tb]
	\caption{\ts($l$)}
	\label{alg:twostage}
	\begin{algorithmic}[1]
	\STATE {\bfseries Input:} matrix $\A$, vector $\b$, sparsity level $k$
	\STATE Initialize $\x^1$
	\FOR{$t=1$ {\bfseries to} $T$}
		\STATE $\topelem_{t+1} \gets$ indices of top $l$ elements of $|\A^T(\A \x^t - \b)|$
		\STATE $\nextJ \gets \curr \cup \topelem_{t+1}$
		\STATE $\z^{t+1}_{\nextJ} \gets \A_{\nextJ} \backslash \b,\ \z^{t+1}_{\bar{J}_{t+1}} \gets \zero$
		\STATE $\y^{t+1} \gets \sproj{k}{\z^{t+1}}$
		\STATE $\next \gets \supp(\y^{t+1})$
		\STATE $\nextiter_\next \gets \Anext \backslash \b,\ \nextiter_{\nextc} \gets \zero$
	\ENDFOR	
	\end{algorithmic}
\end{algorithm}

We now provide a simple analysis for the general two-stage hard thresholding algorithms. We first present a few technical lemmas that we will need for our proof. 

\begin{lemma}
Let $b=A\truevec$, where $I^*=\supp(\truevec)$. Also, let $x=\argmin_{\supp(x)=I}\|Ax-b\|^2$. Then, 
 $$\sqrt{\|(\x-\truevec)_{I\cap I^*}\|^2+\|\x_{I\backslash I^*}\|^2}=\|(\x-\truevec)_{I}\|\leq \frac{\delta_{|I\cup I^*|}}{\sqrt{1-\delta_{|I\cup I^*|}^2}}\|\truevec_{I^*\backslash I}\|$$
\label{lem:errup}
\end{lemma}
\begin{proof}
A similar inequality appears in \cite{Foucart10} and we rewrite the proof here. Since $\x_I$ is the solution to $\min_{u} \|A_Iu-b\|^2$, 
 \begin{equation}
 \A_I^T(A_I\x_I-b)=0.
 \label{eq:ls1}
 \end{equation}
 In the exact case, $b=A\x^*$. 
 Hence, 
 \begin{align}
   \|(\x-\truevec)_{I}\|^2=\left[(\x-\truevec)_{I}\ \ 0\right]^T\left[\begin{matrix}(\x-\truevec)_{I}\\ -\truevec_{I^*\backslash I}\end{matrix}\right]
 \label{eq:t1}
 \end{align}
 Now, using \eqref{eq:ls1}:
 \begin{equation}
 0=\left[(\x-\truevec)_{I}\ \ 0\right]^TA_G^TA_G\left[\begin{matrix}(\x-\truevec)_{I}\\ -\truevec_{I^*\backslash I}\end{matrix}\right],
 \label{eq:t2}
 \end{equation}
 where $G=[I\ \ I^*\backslash I]$.
 Subtracting \eqref{eq:t2} from \eqref{eq:t1} we get,
 \begin{align}
 \|(\x-\truevec)_{I}\|^2&=\left[(\x-\truevec)_{I}\ \ 0\right]^T(I-A_G^TA_G)\left[\begin{matrix}(\x-\truevec)_{I}\\ -\truevec_{I^*\backslash I}\end{matrix}\right],\nonumber\\
 &\leq  \delta_{2k}\|(\x-\truevec)_{I}\|\sqrt{\|(\x-\truevec)_{I}\|^2+\|\truevec_{I^*\backslash I}\|^2},
 \label{eq:t3}
 \end{align}
 where  the second inequality follows using Lemma~\ref{lem:lem1}. Lemma follows by just rearranging terms now. 
\end{proof}

We now present our main theroem and its proof for two-stage thresholding algorithms. 
\begin{theorem}
Suppose the vector $\truevec \in \mathbb{R}^n$ is $k$-sparse and binary. Then \ts($l$)\ recovers
$\truevec$ from measurements $\b = \A \truevec$ in $O(k)$ iterations provided:
$$\delta_{2k+l}\leq .35$$
\end{theorem}
\begin{proof}
As $\z^{t+1}$ is the least squares solution over support set $\nextJ$, hence:
\begin{equation}
f(\z^{t+1})-f(\x^t)\leq f(s^{t+1})-f(\x^t),
\label{eq:fzs}
\end{equation}
where $s^{t+1}_{\nextJ}=(x^t-\eta A^T(Ax^t-b))_{\nextJ}$, $\eta=\frac{1}{1+\delta_l}$ and $s^{t+1}_{\bar{J}_{t+1}}=0$. 

Now, 
\begin{align}
\label{eq:fs}
  f(s^{t+1})-f(\x^t)&=(s^{t+1}-x^t)^TA^T(Ax^t-b)+\frac{1}{2}\|As^{t+1}-Ax^t\|^2. 
\end{align}
Now, as $x^t$ is the least squares solution over $I_t$. Hence, $A_{I_t}^T(Ax^t-b)=0$. 
Hence, 
\begin{equation}
\label{eq:sx}
(s^{t+1}-x^t)_{I_t}=0,\ \ \ (s^{t+1}-x^t)_{\topelem_{t+1}}=-\eta A^T_{\topelem_{t+1}}(Ax^t-b),\ \ \ (s^{t+1}-x^t)_{\bar{J}_{t+1}}=0.
\end{equation}

Using \eqref{eq:fs} and \eqref{eq:sx}:
\begin{align}
 f(s^{t+1})-f(\x^t)&=-\eta\|A^T_{\topelem_{t+1}}(Ax^t-b)\|^2+\frac{\eta^2}{2}\|A_{\topelem_{t+1}}A^T_{\topelem_{t+1}}(Ax^t-b)\|^2,\nonumber\\
&\leq -\eta\|A^T_{\topelem_{t+1}}(Ax^t-b)\|^2+\frac{\eta^2(1+\delta_{l})}{2}\|A^T_{\topelem_{t+1}}(Ax^t-b)\|^2,\nonumber\\
&= -\frac{\eta}{2}\|A^T_{\topelem_{t+1}}(Ax^t-b)\|^2.
\label{eq:fsup}
\end{align}

Now, let $\md$ be the set of missed detections, i.e., $\md=I^*\backslash I^t$. Then, by definition of $\topelem_{t+1}$:
\begin{equation}
\label{eq:topmd}
  \|A^T_{\topelem_{t+1}}(Ax^t-b)\|^2\geq \min\left(1, \frac{l}{|\md|}\right) \|A^T_{\md}(Ax^t-b)\|^2. 
\end{equation}

Furthermore, 
\begin{align}
  \|A^T_{\md}(Ax^t-b)\|&=\|A^T_{\md}A_{\md}\x^*_{\md}-A^T_{\md}A_{I_t}(x^t-x^*)_{I_t}\|,\nonumber\\
&\geq \|A^T_{\md}A_{\md}\x^*_{\md}\|-\|A^T_{\md}A_{I_t}(x^t-x^*)_{I_t}\|,\nonumber\\
&\geq (1-\delta_k)\|\x^*_{\md}\|-\frac{\delta_{2k}^2}{\sqrt{1-\delta_{2k}^2}}\|\x^*_{\md}\|,
\label{eq:mdlow}
\end{align}
where last inequality follows using Lemma~\ref{lem:lem2} and Lemma~\ref{lem:errup}. 

Hence, using \eqref{eq:fsup}, \eqref{eq:topmd}, and \eqref{eq:mdlow}:
\begin{equation}
  \label{eq:fsup1}
  f(z^{t+1})-f(\x^t)\leq f(s^{t+1})-f(\x^t)\leq -\frac{1}{2(1+\delta_l)}\min\left(1, \frac{l}{|\md|}\right)\left(1-\delta_k-\frac{\delta_{2k}^2}{\sqrt{1-\delta_{2k}^2}}\right)^2\|\x^*_{\md}\|^2. 
\end{equation}

Next, we upper bound increase in the objective function by removing $l$ elements from $z^{t+1}$. 
\begin{align}
  f(y^{t+1})-f(z^{t+1})&=(y^{t+1}-z^{t+1})^TA^T(Az^{t+1}-b)+\frac{1}{2}\|Ay^{t+1}-Az^{t+1}\|^2,\nonumber\\
&= \frac{1}{2}\|Ay^{t+1}-Az^{t+1}\|^2,\nonumber\\
&\leq \frac{1+\delta_{l}}{2}\|z^{t+1}_{J_{t+1}\backslash I_{t+1}}\|^2,
  \label{eq:fzy}
\end{align}
where the second equation follows as $z^{t+1}$ is a least squares solution, and both $y^{t+1},\ z^{t+1}$'s support is a subset of $J_{t+1}$. The third equation follows from RIP and the fact that $z^{t+1}_{I_{t+1}}=y^{t+1}_{I_{t+1}}$. 

Now, using Lemma~\ref{lem:errup}:
\begin{equation}
  \label{eq:fzrem}
  \|z^{t+1}_{\nextJ\backslash I^*}\|^2\leq \frac{\delta_{2k+l}^2}{1-\delta_{2k+l}^2}\|\x^*_{I^*\backslash \nextJ}\|^2. 
\end{equation}
Furthermore, $|J_{t+1}\backslash I_{t+1}|=l\leq |\nextJ\backslash I^*|$. Hence, by definition of $I_{t+1}$,
$$\|z^{t+1}_{J_{t+1}\backslash I_{t+1}}\|^2\leq \frac{l}{|\nextJ\backslash I^*|}\|z^{t+1}_{\nextJ\backslash I^*}\|^2.$$
Using above equation and \eqref{eq:fzrem}, we get:
\begin{equation}
  \label{eq:fzup}
\|z^{t+1}_{J_{t+1}\backslash I_{t+1}}\|^2\leq \frac{l}{|\nextJ\backslash I^*|}\frac{\delta_{2k+l}^2}{1-\delta_{2k+l}^2}\|\x^*_{I^*\backslash \nextJ}\|^2,
\end{equation}

Also, $|\nextJ\backslash I^*|=l+|I^*\backslash \nextJ|\leq l+|MD_t|$. 
Using \eqref{eq:fzy}, \eqref{eq:fzup}, and the fact that $f(\x^{t+1})\leq f(y^{t+1})$ and each $x^*_{I^*}=1$:
\begin{equation}
  \label{eq:fxz}
  f(x^{t+1})-f(z^{t+1})\leq \frac{l}{l+|I^*\backslash \nextJ|}\frac{1+\delta_{l}}{2} \frac{\delta_{2k+l}^2}{1-\delta_{2k+l}^2}|I^*\backslash \nextJ|.
\end{equation}

Adding \eqref{eq:fsup1} and \eqref{eq:fxz}, we get:
\begin{equation}
  \label{eq:fxx4}
  f(x^{t+1})-f(x^t)\leq -\frac{1}{2(1+\delta_l)}\left(\min\left(|\md|, l\right)\left(1-\delta_k-\frac{\delta_{2k}^2}{\sqrt{1-\delta_{2k}^2}}\right)^2-\frac{l\cdot |I^*\backslash \nextJ|}{l+|I^*\backslash \nextJ|} \frac{(1+\delta_{l})^2\delta_{2k+l}^2}{1-\delta_{2k+l}^2}\right).
\end{equation}

Now, $\frac{l\cdot |I^*\backslash \nextJ|}{l+|I^*\backslash \nextJ|}\leq \min(l,|I^*\backslash \nextJ|)\leq \min(l,|\md|)$. 

Hence,
\begin{equation}
  \label{eq:fxx}
  f(x^{t+1})-f(x^t)\leq -\frac{\min(l,|\md|)}{2(1+\delta_l)}\left(\left(1-\delta_k-\frac{\delta_{2k}^2}{\sqrt{1-\delta_{2k}^2}}\right)^2- \frac{(1+\delta_{l})^2\delta_{2k+l}^2}{1-\delta_{2k+l}^2}\right).
\end{equation}

Now consider:
\begin{align}
  \left(\left(1-\delta_k-\frac{\delta_{2k}^2}{\sqrt{1-\delta_{2k}^2}}\right)^2- \frac{(1+\delta_{l})^2\delta_{2k+l}^2}{1-\delta_{2k+l}^2}\right)&\geq \frac{1}{1-\delta_{2k+l}^2}\left(((1-\delta_{2k+l})\sqrt{1-\delta_{2k+l}^2}-\delta_{2k+l}^2)^2-(1+\delta_{2k+l})^2\delta_{2k+l}^2\right),\nonumber\\
&>0.01,
\label{eq:fxx1}
\end{align}
where the second inequality follows by substituting $\delta_{2k+1}\leq .35$. 

Hence, using \eqref{eq:fxx} and \eqref{eq:fxx1}, we have:
\begin{equation}
  \label{eq:fxx2}
  f(x^{t+1})\leq f(x^t)-\min(l,|\md|)\cdot 0.0001. 
\end{equation}
The above equation guarantees convergence to the optima in at least $O(k)$ steps although faster convergence can be shown for larger $k$. 
\end{proof}

\begin{corollary}
  Cosamp converges to the optima provided $$\delta_{4k}\leq 0.35.$$
\end{corollary}
\begin{corollary}
  Subspace-Pursuit  converges to the optima provided $$\delta_{3k}\leq 0.35.$$
\end{corollary}

Note that CoSamp's analysis given by \cite{NeedellT09} requires $\delta_{4k}\leq 0.1$ while Subspace pursuit's analysis given by \cite{DaiM09} requires $\delta_{3k}\leq 0.205$. Note that our generic analysis provides significantly better guarantees for both the methods. 


\end{document}